\documentstyle[12pt,epsf]{article}
\newcommand{\U}{{\cal U}}
\newcommand{\beq}{\begin{equation}}
\newcommand{\eeq}{\end{equation}}
\newcommand{\beqa}{\begin{eqnarray}}
\newcommand{\eeqa}{\end{eqnarray}}

\newcommand{\R}{{\bf R}}
\newcommand{\C}{{\bf C}}
\newcommand{\CP}{{\bf P}}

\begin{document}

\begin{titlepage}

\begin{center}
\today
\hfill HUTP-97/A024, LBNL-40349, UCB-PTH-97/27\\
\hfill                  hep-th/9705220

\vskip 2 cm
{\Large\bf
Non-Abelian Conifold Transitions\\[0.3cm]
and
$N=4$ Dualities in Three Dimensions}
\vskip 1.5 cm
{Kentaro Hori$^{1,2}$, Hirosi Ooguri$^{1,2}$ and Cumrun Vafa$^3$}\\
\vskip 0.8cm
{\sl $^1$ Department of Physics,
University of California at Berkeley\\
366 Le\thinspace Conte Hall, Berkeley, CA 94720-7300, USA\\}
\vskip .5cm
{\sl $^2$ Theoretical Physics Group, Mail Stop 50A--5101\\
Ernest Orlando Lawrence Berkeley National Laboratory\\ Berkeley, CA
94720, USA\\}
\vskip .5cm
{\sl $^3$ Lyman Laboratory of Physics, Harvard University\\
Cambridge, MA 02138, USA}

\end{center}

\vskip 0.5 cm
\begin{abstract}
We show how Higgs mechanism for non-abelian $N=2$ gauge theories
in four dimensions
is geometrically realized in the context of type II strings
as transitions among compactifications of Calabi-Yau
threefolds.  We use this result and T-duality of a further
compacitification on a circle to derive $N=4$, $d=3$ dual
field theories.  This reduces dualities
for $N=4$ gauge systems in three dimensions to
perturbative symmetries of string theory.  Moreover
we find that the dual of a gauge system always exists
but may or may not correspond to a lagrangian system.  In particular
we verify a conjecture of Intriligator and Seiberg
that an ordinary gauge system is dual to compacitification
of Exceptional tensionless string theory down to three dimensions.
\end{abstract}

\end{titlepage}

\baselineskip=0.7cm

\section{Introduction}

One of the most important lessons we have learned
recently in string theory is the fact that interesting
field theories
can be realized by considering singular compactifications
of string theory with or without D-branes
present.  In this setup one can translate
aspects of field theories in question to facts
about the geometry of the manifold.
This general idea is known as {\it geometric engineering}.

One of the main powers of geometric engineering is the flexibility
in constructing any field theories we wish to construct.  This
is perhaps the most important aspect of this method (for example
the construction of exceptional gauge groups has not been done
in a geometrically faithful way in any other approach).
But in addition, and what seems
to be very surprising at first sight, is that in this setup the
non-trivial field theory dualities can in one way or another
be reduced to {\it classical symmetries} of string theory.
  This seems quite surprising.
This can be done in particular for $N=4$ theories in $d=4$
by considering type IIA on ALE space of ADE type times $T^2$, where
T-duality of $T^2$ is a geometric realization of Olive-Montonen
S-duality  for the ADE group \cite{wit, duff}.
Similarly exact results for $N=2$ gauge systems
can be obtained by
geometric engineering of type II strings on Calabi-Yau
threefolds \cite{kkv, kmv},
and by using mirror symmetry which is a classical symmetry
of type II strings.  This approach has been extended to
$N=1$ theories in
$d=4$ in \cite{kv, bereta,
ov, Ah, Ano} in which the
dualities are realized as classical symmetries of strings.
Similarly higher
dimensional critical theories (with tensionless
strings) have also been constructed
from this viewpoint and in particular
$N=1$ theories in five dimensions
\cite{dkv, ms, mis} and $N=1$
theories in six dimensions
\cite{witsix, bv, i, ma}
have been engineered.
In certain cases
 constructions can also be done using D-branes in the presence of
NS 5-branes \cite{hw, dhooz,
kut, more1, more2, more3, more4, more5,
more6, more7, more8},
and often there is a T-duality \cite{OV1}
which connects the two pictures (see in particular \cite{ov}).

An interesting duality was proposed for
three dimensional theories with $N=4$ in \cite{intse}.
This was further extended to a large number of non-abelian
gauge theories in \cite{dhoo, PZ}.
So far, the only approach from string theory
involving a derivation
of $N=4$ dualities in $d=3$ with non-abelian gauge groups
involves the use
of non-perturbative string dualities \cite{hw, dhooz}.
One of our aims in this paper is to show how duality
 of $N=4$ theories in $d=3$ can also be reduced
to classical symmetries of type II strings.  This is
done by constructing local models of $N=4$ gauge systems
involving a non-compact Calabi-Yau threefold times a circle
and using the T-duality of the circle to exchange type IIA and type
IIB strings.  The main ingredient needed in this description is a precise
understanding of how the Coulomb/Higgs phases of the gauge system
are realized geometrically.  Realization of Coulomb branches have been
understood in the type IIA \cite{klm, kamp,
geosin, mge, morerefs},
and the type IIB setup \cite{bsav, klmvw, kkv, kmv}.
  However much less is known
about the Higgs branch.  In this paper we will develop techniques to
describe the Higgs branch in a geometrical way.

This construction not only allows us to rederive
the $N=4$ dualities in $d=3$ from perturbative symmetries
of strings, but it also allows us to see why in some
cases the dual of a gauge system is {\it not} a lagrangian quantum
field theory.  A special case of this was already
conjectured in \cite{intse}, which we shall verify
in this paper.  We believe this is actually
an important lesson, far more general than the example being
studied here.  In particular if we wish to find
dual pairs for all field theories we should broaden the
class of field theories under study to include non-lagrangian
quantum field theories, which have been encountered
in string theory (and the higher dimensional versions of
which are distinguished by the appearance of tensionless strings).
This may also explain why the search for dual pairs of gauge theories
in four dimensions
have been incomplete so far.  In fact based on
the three dimensional theories we study in this paper
it is natural to  conjecture that
{\it for every quantum field theory in any dimension
there are dual descriptions, which
may or may not involve lagrangian systems}.  We can verify this
conjecture for the $N=4$ theories in $d=3$
which can be geometrically engineered.  In this case
the existence of a dual description is an automatic
consequence of our setup.

The organization of this paper is as follows:  In section
2 we introduce the basic idea and review some facts about
$N=4$ dualities in $d=3$.  In section 3, in anticipation
of applications in section 4, we review the resolution
of ADE singularities of ALE spaces in detail (which
is self-contained and
we hope is accessible to the reader).  In section 4
we show how $N=2$ Higgs mechanism  in four dimensions
is related to the resolution
of certain singularities in type IIB string context
and use this result to derive $N=4$, $d=3$ dual pairs.
Also in this section we discuss the dual of toroidal
compactifications of Exceptional tensionless strings down
to three dimensions.

\section{Basic Idea}

We consider compactifications of type
IIA and IIB strings on  Calabi-Yau 3-folds. In such a compactification
 we generically obtain
an effective $N=2$, $d=4$ theory with some
number of $U(1)$'s, in which the vector multiplet
moduli space ({\it Coulomb branch}) of the theory gets identified
with the complex/K\"ahler moduli of Calabi-Yau
and the hypermultiplet moduli space ({\it Higgs branch})
gets identified with the Jacobian variety over K\"ahler/complex
moduli of Calabi-Yau in the type IIB/A respectively.
In the latter case, we consider the Jacobian in order to take
into account the RR field configurations on the Calabi-Yau.\footnote{
In four dimensions the vector multiplets do not receive any
quantum string corrections whereas the hypermultiplet moduli do.
However if we go down to three dimensions
on a further circle Coulomb branch also receives quantum corrections.}

Depending on whether we put type IIA or IIB on a fixed 3-fold,
in general we get inequivalent theories in 4 dimensions.
However upon further compactification on $S^1$,
they become equivalent by T-duality on $S^1$. The effective
$N=4$, $d=3$ theories are therefore also equivalent, but their
Coulomb and Higgs branches are exchanged. In fact, such an
exchange symmetry in $N=4$, $d=3$ gauge theories were found in \cite{intse}
and was called the mirror symmetry in 3 dimensions.  That it should
be a consequence of the T-duality of the type IIA and IIB theories
was suggested in \cite{sesh}. In principle, this should
explain all the mirror symmetries of $N=4$, $d=3$ gauge theories
which arise from type II theory on a Calabi-Yau 3-fold. In fact,
in \cite{dhoo, dhooz}, it was shown in detail how it works when
the gauge group is a product of $U(1)$'s.
In practice, however, it is difficult
to apply this idea directly in non-abelian cases. This is because,
as we will see below, we would need
to find non-abelian generalization of the conifold transition
\cite{str}, \cite{gms}. It turns out that the task is significantly
simplified if we use mirror symmetry for Calabi-Yau
threefolds. Let us describe our
strategy to analyze the non-abelian case by first reviewing the
abelian case.

\subsection{Duality in the Abelian Case}

As is well known, D-branes wrapped around cycles of Calabi-Yau
give rise to solitons in this geometry.  In particular if
one considers type IIB with an $S^3$ inside a Calabi-Yau threefold $W$,
by wrapping a $D3$ brane around $S^3$ we obtain a charged hypermultiplet
\cite{str}
(charged under the $U(1)$ obtained by decomposition of the 4-form
RR gauge potential as the volume form on $S^3$ times a gauge field
in space time).  Moreover the mass of the hypermultiplet
is proportional to the volume of $S^3$, which thus vanishes
in the limit $S^3$ vanishes. This vanishing can be accomplished
by changing the complex structure of Calabi-Yau.
 If we have more vanishing $S^3$'s
than the number of $U(1)$'s then we can consider higgsing the $U(1)$'s.
In particular it was shown in \cite{gms}\
(see also \cite{gmv}) that
this leads to a transition to a new Calabi-Yau in which
we blow up some $S^2$'s,
as anticipated in \cite{canHu}.
Let us denote the Calabi-Yau we started with by $W$
and the one we obtain after transition by $ W^t$.
In geometrical terms, we have tuned the complex moduli
of $ W$ to get a singular Calabi-Yau and then
changed the K\"ahler structure of the singular space
to obtain $ W^t$ after transition to the Higgs branch.
To be concrete let us assume that we Higgs a $U(1)^k$ system
with $N>k$ hypermultiplets.  Let $h^{p,q}$  denote the Hodge
number of Calabi-Yau. Then we have
\beqa
&h^{2,1}(W)-k = h^{2,1}({W}^t) \\
&h^{1,1} (W) = h^{1,1}(W^t)-(N-k)
\eeqa
If we consider type IIA instead of type IIB,
we have an interpretation of the same transition
in terms of a (generically)
inequivalent theory in 4 dimensions.  In particular
the inverse
of the transition, namely  $W^t\rightarrow W$ will
have the interpretation of the Higgsing of $U(1)^{N-k}$ with
$N$ flavors.  Note in particular that what appears in the type IIB
as the Higgs branch is now related to the Coulomb branch of a
type IIA theory.

In 4 dimensions these two theories are inequivalent.
However if we compactify the theories on an extra circle
the story changes.  This is because T-duality on the circle
relates type IIA on ${W}\times S^1$ to type IIB on
${W}\times S^1$. Thus when we take the circles
to be of the order of the string scale, after decoupling
the excited modes of string,
we obtain two effective 3-dimensional theories which should be equivalent.
This duality of field theories in $3d$ is known as mirror symmetry
\cite{intse}, and the connection to the above transition in Calabi-Yau
was noted in \cite{dhoo}.

This duality symmetry, which we discussed
in the abelian case above, has been extended to
non-abelian gauge groups  \cite{dhoo, hw,
dhooz}. We would like to find the non-abelian
realization of these transitions
in Calabi-Yau compactifications
in the same way we did for the abelian case above, and
thus reduce the $N=4$, $d=3$ dualities to a perturbative symmetry
of string theory.

\subsection{Generalization to Non-Abelian Case}

It is natural to expect that the derivation
of the abelian duality symmetry in three dimensions involving the transition
of Calabi-Yau will have non-abelian generalization. Finding
this generalization will be useful as will reduce 3d duality
symmetry of $N=4$ theories to the
 knowledge available from string perturbation theory, plus
physical interpretation of extremal transitions among Calabi-Yau
in terms of Higgs/Coulomb branch transitions,
which is more or less understood in terms
of the D-brane solitons.

Mirror symmetry of Calabi-Yau is very important in understanding
the non-abelian case.
 Let $M$ denote a
Calabi-Yau threefold and $ W$ be its mirror.  By
definition, this means that type IIA/B on $ M$ gives
the same theory as type IIB/A on $ W$, where the role
of complex deformations and K\"ahler deformations get exchanged.
By now there is a lot of evidence for this symmetry \cite{yau}
and some of it has been rigorized
\cite{kont, give}.  Moreover there are
hints that this symmetry is related to the more familiar
$T$-duality ($R\rightarrow 1/R$) symmetry of toroidal compactification
where one views the threefold as a $T^3$ fibered over $S^3$
\cite{syz, mor, gross} (see
also \cite{vw}.)
 We will consider type IIA
on a local model of Calabi-Yau 3-fold, $M_K$
whose K\"ahler deformations
give the Coulomb branch of an $N=2$ theory in $d=4$ (the
subscript $K$ is there to remind us that we are considering
varying the K\"ahler structure).  We also consider the completely Higgsed
branch which corresponds to an extremal transition of $M_K\rightarrow
M^t_C$.  The subscript $C$ on $M^t_C$ is to remind
us that the complex structure variation corresponds to the Higgs
branch of the theory.  We thus consider
\beq
IIA(M_K,M^t_C)
\eeq
as a local model for the Coulomb and Higgs branch of an $N=2$
theory in $d=4$ in the context of type IIA strings. Using mirror
symmetry the same theory can be described equivalently as
\beq
IIA(M_K,M^t_C)=IIB(W_C,W^t_K)
\eeq
Now we consider compactifying on the circle
to get an $N=4$ theory in $d=3$. We thus have
\beq
IIA((M_K,M^t_C)\times S^1)=IIB((W_C,W^t_K)\times
S^1)=IIA((W_C,W^t_K)\times {\hat S}^1)
\eeq
where ${\hat S}^1$ denotes the T-dual circle.  We thus conclude
two type IIA models with (Coulomb,Higgs) branches given by
the local model $(M_K,M^t_C)$ and $(W^t_K,W_C)$ which are
inequivalent theories in 4 dimensions, will become equivalent
in 3 dimensions, where the role of K\"ahler and complex deformations
are exchanged.

This is a general correspondence between
two theories in $3d$ and {\it it holds whether or not
the local model of Calabi-Yau's correspond to
any gauge systems}.  In case that both the $M$ and $W$
lead to identifiable gauge systems we can
then deduce dual gauge systems in $3d$.  If one of them
is a gauge system and the other is not we learn that a 3d $N=4$
field theory may have a dual which is {\it not} a gauge system.
Some cases of this type were conjectured in \cite{intse} and we will
actually verify their conjecture.  There are in principle
also cases where neither side is a gauge system, and
we would have a 3d field theory duality not
involving gauge systems.  We shall
not consider this last case in this paper (but it may very
well be the generic case).

For the purpose of identifying the 3d mirror
for a gauge system we would need to know the local
model for the Calabi-Yau 3-fold corresponding to a given
group and matter.  This can be done by geometric engineering
of quantum field theory \cite{kkv}.
In particular if we are interested
in pure $N=2$ gauge system, in type IIA compactification
we need to fiber an A-D-E
singularity over $P^1$. The K\"ahler parameters corresponding
to the blowing up of A-D-E singularity will correspond to Coulomb
moduli of the corresponding gauge system.  If we wish to get
matter we will obtain it by ``colliding singularities'' which means
that we consider
 intersecting $P^1$'s over which we have
A-D-E singularities.  Depending on what singularity
is on top of intersecting $P^1$'s we will get matter
in various representations \cite{geosin,
mge}\footnote{Not all intersecting
singularities will give rise to matter, for example two $D$-type
singularities
intersecting gives rise to a superconformal $N=2$ system
which has no matter interpretation\cite{bj}.}.
For example if we wish to get $U(n) \times U(m)$ with matter in
bi-fundamental $(n,{\overline m})$ we consider a type IIA geometry with
two intersecting $P^1$'s over one having an $A_{n-1}$ singularity and over
the other an $A_{m-1}$ singularity and at the intersection
point an $A_{n+m-1}$ singularity. The bi-fundamental $(n,{\overline m})$
can be interpreted as part of the decomposition of the
adjoint matter of $U(n+m)$ and thus the corresponding bi-fundamental
matter is localized near the intersection point \cite{geosin,
mge, mw}. In general if we have enough
matter we can also Higgs the system and consider the complex
deformations of the manifold which correspond to the Higgs branch.
This in particular means that we go (classically) to the
origin of Coulomb branch, i.e. blow down the A-D-E
fibers and then deform the singularity of the manifold
by changing the complex structure.  For simple gauge systems
such as $U(n)$ with fundamentals such a description
is possible and is known \cite{geosin}, \cite{mge}, but for more complicated
systems it has not been worked out (our results below will
amount to a description of this for a large number of cases).
However what is known, in some simple
cases
\cite{kkv}\ and in much greater generality in \cite{kmv},
is how to describe the exact Coulomb branch
of gauge systems which are geometrically constructed
in type IIA setup by applying local mirror symmetry
and converting it to a type IIB compactification.

So our basic strategy is to start with a group $G$ (not necessarily
simple)
and representation $R$
and consider the local 3-fold $M_K$ which in type IIA gives rise to it.
Then use the results \cite{kmv} to construct the mirror manifold
in type IIB, which we denote by $W_C$.
Then we will explicitly construct what complete Higgsing means in this setup
by finding $W^t_K$ and then consider what matter structure
 $W^t_K$ would correspond to if it were viewed in type IIA context.
As a passing remark note
 that this also gives the $M^t_C$ manifold by applying mirror symmetry
to $W^t_K$ (using the results in \cite{kmv}),
i.e. we will be able
to write down the geometry corresponding to the non-abelian
Higgs phenomenon.

\subsection{Examples of Dual Pairs}

Let us review some of the known examples of dual pairs
of $N=4$ gauge theories
in three dimensions. Examples with $U(1)$ and $SU(2)=Sp(1)$ gauge
groups were studied in \cite{intse}, and they were
generalized cases with higher rank gauge groups in \cite{dhoo, PZ}.
In these models, hypermultiplet moduli spaces
do not receive any quantum corrections and can be read off directly from
their classical Lagrangians by the hyperk\"ahler quotient
construction. On the other hand, their vector multiplet moduli
spaces may be deformed by quantum effects.

\vskip .1in

\noindent
{\it Example 1}

\vskip .05in

\noindent
A-model: $U(k)$ gauge group with $1$ adjoint and
$n$-fundamental matters

\noindent
B-model: $\prod_{i=1}^n U(k)_i$ gauge group with a fundamental matter
in $U(k)_1$. The is also a bi-fundamental for each
$U(k)_i \times U(k)_{i+1}$
where $i=1,...,n$ and
$U(k)_{n+1} = U(k)_1$,
\vskip .1in

The maximum Higgs branch of the A-model is the moduli space of
$SU(n)$ instantons of degree $k$,
and the maximum Higgs branch of the B-model is the Hilbert scheme
(resolved symmetric product) of $k$-points on the $A_n$-type
ALE space. There
are also various mixed branches of these models. In \cite{dhoo},
it is shown how these branches
transform into each other under the duality, by taking
into account quantum corrections to the vector multiplet
moduli spaces. The duality transformation,
 which exchanges the mass and the FI parameters of the two
models, was also found.

\vskip .1in

\noindent
{\it Example 1'}

\vskip .05in

It is possible to eliminate the adjoint matter in the A-model
by adding its mass term. According to the mirror map, this
corresponds to turning on the FI parameter for the diagonal
$U(1)$ of $U(k)^n$. We then obtain

\vskip .05in

\noindent
A-model: $U(k)$ gauge group with $n$-flavors.

\noindent
B-model: $\prod_{i=1}^{n-1} U(l_i)$
gauge group with
\beq
(l_1 , l_2, ..., l_n) =
(1, 2,...(k-1), k, k, ..., k, (k-1), ... , 2, 1)
\eeq
($l_i=k$ for $i=k,...,(n-k)$).
There is a fundamental for $U(l_k)$ and $U(l_{n-k})$,
and a bi-fundamental
for each $U(l_i) \times U(l_{i+1})$.

\vskip .1in

These models can be generalized to include
arbitrary linear chain of $U(k_i)$ groups with
bifundamental matter as well as possible extra fundamental
matter and their dual turns out also to be of the same type
(and it is easily derivable from case 4 below).
Hanany and Witten \cite{hw} pointed out that these
models can be constructed by using webs of
NS 5-branes and D3 and D5-branes in type IIB string theory,
and suggested that the duality in this case
is a consequence of the $SL(2,Z)$ $S$-duality of the
type IIB theory.

In the Abelian ($k=1$) case, this example
reduces to the $A_n$ type dual pairs
of \cite{intse}. In this case, it was pointed out in
\cite{dhoo} that the mirror symmetry is a consequence of
the $T$-duality of the type IIA and IIB theories. In the following,
we will see how this observation is generalized to the non-Abelian
($k>1$) case.

\vskip .1in

\noindent
{\it Example 2}

\vskip .05in

\noindent
A-model: $Sp(k)$ gauge group with one antisymmetric representation
and and $n$-fundamental matters.

\noindent
B-model: $\prod_{i=1}^{n-3} U(2k)_i \times
\prod_{i=1}^4 U(k)_i$. There is a fundamental
for $U(k)_1$. There is also a bi-fundamental
for each $U(2k)_i \times U(2k)_{i+1}$
($i=1,...,n-4$) and also for
$U(2k)_1 \times U(k)_1$, $U(2k)_1 \times U(k)_2$,
$U(2k)_{n-3} \times U(k)_3$ and $U(2k)_{n-3} \times U(k)_4$.

\vskip .1in

In this case, the maximum Higgs branch of the A-model is
the moduli space of $SO(n)$ instantons of degree $k$,
and the one for the B-model is
the Hilbert scheme of $k$ points on the $D_n$-type ALE space.

\vskip .1in
\noindent
{\it Example 2'}

\vskip .05in

As in the case of example 1', we can turn on the
mass parameter for the matter in the antisymmetric representation
in the A-model and the corresponding FI parameter for the
B-model. The resulting mirror pair is:

\vskip .05in

\noindent
A-model: $Sp(k)$ gauge group with $n$-fundamentals.

\noindent
B-model: $\left[\prod_{i=1}^{n-2} U(l_i)\right] \times U(k)_1 \times
U(k)_2$ gauge group with
\beq
 (l_1,...,l_{n-2}) = (1,2,...,(2k-1),2k,....,2k)
\eeq
($l_i=2k$ for $i=2k,...,n-2$).
There is a fundamental in $U(l_{2k})$.
there is also a bi-fundamental for each
$U(l_i) \times U(l_{i+1})$, ($i=1,...,{n-3}$),
and also for $U(l_{n-2}) \times U(k)_1$ and $U(l_{n-2})
\times U(k)_2$.

\vskip .1in
The Abelian case ($k=1$) corresponds to the $D_n$-type dual pair in
\cite{intse}.  We will verify this duality for general $k$
in this paper.

\vskip .1in
\noindent
{\it Example 3}

\vskip .05in
In \cite{intse}, it was conjectured that if we consider a model
whose gauge group is a product of $U(l_i)$'s arranged
on nodes of the affine $E_n$ ($n=6, 7, 8$) Dynkin diagram
with $l_i$ being equal to the Dynkin index of each node,
its Coulomb branch is the moduli space of $E_n$ instantons of degree $1$,
and that it is dual to the compactification of tensionless $E_n$ string
theories to three dimensions.  We will verify this conjecture in this
paper.  In \cite{kmv} the Coulomb branch for product
of $U(kl_i)$ gauge groups arranged on nodes of the affine $E_n$
Dynkin diagram has been found, where $k$ is an arbitrary integer
and $l_i$ is the Dynkin index of the corresponding node.  It is shown
there, using this result, that this system is dual to $k$ small
$E_n$ instantons compactified to $d=3$, as conjectured in \cite{intse},
thus extending what we have found here for $k=1$ to higher $k$'s.

\vskip .1in
\noindent
{\it Example 4}

\vskip .05in
The example 1 can be further generalized \cite{dhoo} as

\vskip .05in
\noindent
A-model: $\prod_{i=1}^n U(k)_i$ with $v_i$ fundamental matters in
$U(k)_i$ and a bi-fundamental
for $U(k)_i \times U(k)_{i+1}$ ($i=1,...,n$).

\noindent
B-model: $\prod_{i=1}^m U(k)_i$ with $w_i$ fundamental matters in
$U(k)_i$ and a bi-fundamental
for $U(k)_i \times U(k)_{i+1}$ ($i=1,...,w$).

\vskip .1in
They make a mirror pair if a Young diagram with rows of lengths
$v_1,...,v_n$ is related to a diagram with rows of lengths
$w_1,...,w_m$ by transposition. This, in particular, means
$n=\sum_i w_i$ and $m=\sum_i v_i$. It was pointed out in
\cite{dhooz} that one can construct these models as webs
of NS 5-branes, D3 and D5-branes, as in \cite{hw},
and that the mirror symmetry follows from the $S$-duality
of the type IIB theory.  We expect that the methods
of this paper (and \cite{kmv}) can be generalized
to also include this case as well as the example 1, thus
covering all the cases conjectured.

\section{Resolution of ADE Singularity}

\bigskip
As discussed in the previous section we need to develop
what Higgsing means geometrically and in particular
in the context of type IIB compactifications on Calabi-Yau
threefolds.  Already in the abelian case, it is clear that
one needs to go to a point on the complex moduli of type IIB
side where there is a singularity (where some 3-cycles shrink)
and blowup instead some 2-cycles at these points.
It is thus not surprising that the non-abelian
generalization would in particular involve understanding blowups
and as it turns out of the singularities of A-D-E type for ALE spaces.
In this section,
we thus give a systematic description of the resolution of the
A-D-E singularities that will be used in the next section.
The equations
\beqa
A_{n-1}&:& \quad xy=z^n\label{A}\\
D_n&:&\quad x^2+y^2z=z^{n-1}\label{D}\\
E_6&:&\quad x^2+y^3+z^4=0\label{E6}\\
E_7&:&\quad x^2+y^3+yz^3=0\label{E7}\\
E_8&:&\quad x^2+y^3+z^5=0\label{E8}
\eeqa
describe complex surfaces embedded
in the affine space $\C^3$ with coordinates $x,y,z$.
Each of them has a singularity at $x=y=z=0$
(we assume that $n\geq 2$ for $A_{n-1}$ and $n\geq 3$ for $D_n$)
and its resolution
means a smooth surface which is mapped to it in such a way that
the map is an isomorphism except at the inverse image of the singular
point $x=y=z=0$. The resolution we are going to describe is the so called
minimal resolution and it turns out that the inverse image of the point
$x=y=z=0$ consists of rational curves (i.e. $\CP^1$'s) whose
intersection matrix is the same as the Cartan matrix of the Lie algebra
indicated by the name of its singularity type.

The resolution is carried out by sequential blow-ups of the ambient space
$\C^3$ at the singular points of the surface.
For $A_{n-1}$ and $D_n$ cases,
this can be done more easily by sequential
blow-ups of planes transversal to lines passing through
the singular points.

\bigskip
\medskip
\noindent
\subsection{Resolution of $A_{n-1}$ singularity}

\bigskip
We can resolve the $A_{n-1}$ singularity $xy=z^n$
by a sequence of blow-ups of complex planes.
We first resolve the simplest $A_1$ singularity.
Let us blow up the $x$-$y$-$z$ space at $x=z=0$.
Namely, we replace the $x$-$y$-$z$ space
by a union of two spaces --- coordinatized by
$(x,y,\widetilde{z})$
and $(\widetilde{x},y,z)$ ---
which are mapped to the $x$-$y$-$z$ space by
$(x,y,z)=(x,y,x\widetilde{z})=(z\widetilde{x},y,z)$.
The $x$-$y$-$\widetilde{z}$ and the $\widetilde{x}$-$y$-$z$
spaces are glued by $\widetilde{z}\widetilde{x}=1$ and
$z=x\widetilde{z}$.
The equation $xy=z^2$ of the $A_1$ singularity looks as
$x(y-x\widetilde{z}^2)=0$ in the $x$-$y$-$\widetilde{z}$ space
and $z(\widetilde{x}y-z)=0$ in the $\widetilde{x}$-$y$-$z$ space.
If we ignore the piece described by
$x=0$ and $z=0$
which is mapped to the $y$-axis $x=z=0$, we obtain a union of
two smooth surfaces --- $U_1=\{y=x\widetilde{z}^2\}$
in the $x$-$y$-$\widetilde{z}$ space
and $U_2=\{\widetilde{x}y=z\}$ in the $\widetilde{x}$-$y$-$z$ space.
The surfaces $U_1$ and $U_2$ are coordinatized
by $(x,\widetilde{z})$ and $(\widetilde{x},y)$ respectively
and are glued together by
$\widetilde{z}\widetilde{x}=1$ and $x\widetilde{z}=\widetilde{x}y$.
Thus, we obtain a {\it smooth} surface.
This is the resolution of the $A_1$ singularity.
This surface is mapped subjectively
onto the original singular $A_1$ surface $xy=z^2$:
$(x,y,z)=(x,x\widetilde{z}^2,x\widetilde{z})$ on $U_1$ and
$(x,y,z)=(\widetilde{x}^2y,y,\widetilde{x}y)$ on $U_2$.
The inverse image of the singular point $x=y=z=0$
is described by $x=0$ in $U_1$ and by $y=0$ in $U_2$.
It is coordinatized by $\widetilde{z}$ and $\widetilde{x}$
which are related by $\widetilde{z}\widetilde{x}=1$, and thus is a
projective line $\CP^1$.

If we started with higher $A_{n-1}$ singularity, the equation
$xy=z^n$ looks as
$y=x^{n-1}\widetilde{z}^n$ in the $x$-$y$-$\widetilde{z}$ space
and $\widetilde{x}y=z^{n-1}$ in the $\widetilde{x}$-$y$-$z$ space
(ignoring the trivial piece $x=0$ and $z=0$).
It is smooth in the $x$-$y$-$\widetilde{z}$ plane
but the part is the $\widetilde{x}$-$y$-$z$
has the $A_{n-2}$ singularity
at $\widetilde{x}=y=z=0$.
Thus, the surface is not yet resolved but it has become
{\it less singular} : $n$ has decreased by one. We can further
decrease $n-1$ by one by blowing up the $\widetilde{x}$-$z$ plane at
$\widetilde{x}=z=0$. Iterating this process, we can finally resolve
the singular $A_{n-1}$ surface. It is straightforward to see that
the resolved space is covered by $n$ planes $U_1$, $U_2$, $U_3$, ..., $U_n$
with coordinates $(x_1,z_1)=(x,\widetilde{z})$,
$(x_2=\widetilde{x},z_2)$, $(x_3,z_3)$, ..., $(x_n,z_n=y)$ which are
mapped to the singular $A_{n-1}$ surface by
\beq
U_i\ni(x_i,z_i)\longmapsto\left\{
\begin{array}{l}
x=x_i^iz_i^{i-1}\\[0.2cm]
y=x_i^{n-i}z_i^{n+1-i}\\[0.2cm]
z=x_iz_i
\end{array}
\right.
\label{projA}
\eeq
The planes $U_i$ are glued together by $z_{i}x_{i+1}=1$ and
$x_{i}z_{i}=x_{i+1}z_{i+1}$.
The map onto the singular $A_{n-1}$ surface is isomorphic except
at the inverse image of the singular point $x=y=z=0$.
The inverse image consists of $n-1$ $\CP^1$s $C_1$, $C_2$, ...,
$C_{n-1}$ where $C_i$ is the locus of $x_{i}=0$ in $U_i$
and $z_{i+1}=0$ in $U_{i+1}$, and is coordinatized by $z_i$ and
$x_{i+1}$ that are related by $z_ix_{i+1}=1$.
$C_i$ and $C_j$ do not intersect unless $j=i\pm 1$,
and $C_{i-1}$ and $C_i$ intersect transversely at $x_i=z_i=0$.
It is also possible to show that the self-intersection of $C_i$
is $-2$. Thus, we see that the intersection matrix of the components
$C_1,\ldots,C_{n-1}$ is the same as the $A_{n-1}$ Cartan matrix.

\begin{figure}[htb]
\begin{center}
\epsfxsize=4.5in\leavevmode\epsfbox{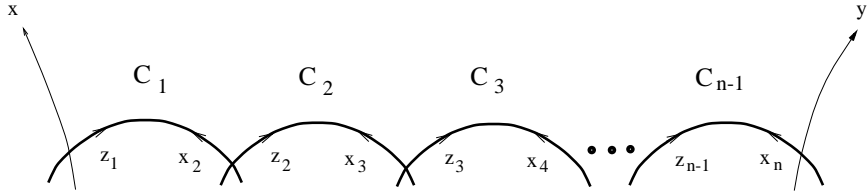}
\end{center}
\caption{resolution of $A_{n-1}$ singularity}
\end{figure}

\bigskip
\noindent
\subsection{Resolution of $D_n$ singularity}

\bigskip
Resolution of $D_n$ singularity ($n\geq 3$) is similar.
Let us first blow up $x=z=0$ and look at the equation
$x^2+y^2z=z^{n-1}$ in
the $x$-$y$-$\widetilde{z}$ space and in the
$\widetilde{x}$-$y$-$z$ space.
Ignoring the trivial piece given by $x=0$ and
$z=0$ in the first and second patches respectively, we see
$x^2+y^2\widetilde{z}=x^{n-2}\widetilde{z}^{n-1}$ in the
$x$-$y$-$\widetilde{z}$ space
and
\beq
z\widetilde{x}^2+y^2=z^{n-2}
\label{Dn-1}
\eeq
in the $\widetilde{x}$-$y$-$z$. Let us assume $n> 3$ for the moment.
Then, the surface is smooth in the $x$-$y$-$\widetilde{z}$ space, but
the part in the $\widetilde{x}$-$y$-$z$ space (\ref{Dn-1})
has a $D_{n-1}$ singularity at the origin $\widetilde{x}=y=z=0$.
We can make it less singular by blowing up the $y$-$z$ plane at
$y=z=0$.
Iterating this process, we finally obtain a $D_3$ singularity.
Now let us consider the $n=3$ case. After blowing up $x=z=0$,
we see $x+y^2\widetilde{z}=x\widetilde{z}^2$
in the $x$-$y$-$\widetilde{z}$ space and
$z\widetilde{x}^2+y^2=z$ in the $\widetilde{x}$-$y$-$z$.
Then, we see that there are two $A_1$ singularities at
$x=y=\widetilde{z}\mp 1=0$
(or equivalently $\widetilde{x}\mp 1=y=z=0$).
Blowing up again at $x=y=0$, we can resolve these $A_1$ singularities
at the same time. In this way, we can resolve the singular
$D_n$ surface.

For later use, we give an explicit description of the resolved
surface.
After the sequence of blow-ups, we obtain a 3-fold covered by
$n$ open subsets $\U_1$, $\U_2,$ $\ldots,$ $\U_n$ with coordinates
$(s_1,t_1,z_1)=(x,y,\widetilde{z})$,
$(s_2=y,t_2=\widetilde{x},z_2),$ $\ldots,$ $(s_n,t_n,z_n)$.
These open sets are glued together
by certain transition relations.\footnote{
For reference, we record the relations:
$(s_j,t_j,z_j)=(s_{j+1}t_{j+1}z_{j+1},s_{j+1},t_{j+1}^{-1})$
for $j=1,\ldots,n-4$,
$(s_{n-3},t_{n-3},z_{n-3})
=(s_{n-2}t_{n-2}^2z_{n-2},s_{n-2}t_{n-2},t_{n-2}^{-1})$,
and
$(s_{n-2},t_{n-2},z_{n-2})
=(z_{n-1}t_{n-1},s_{n-1},t_{n-1}^{-1})
=(t_n^{-1},s_nt_n,z_{n})$.}
The projection to
the $x$-$y$-$z$ space is given by
\beq
\left\{
\begin{array}{lclcl}
x&=&s_{2j-1}^jz_{2j-1}^{j-1}&=&
s_{2j}^jt_{2j}z_{2j}^j\\[0.2cm]
y&=&s_{2j-1}^{j-1}t_{2j-1}z_{2j-1}^{j-1}&
=&s_{2j}^jz_{2j}^{j-1}\\[0.2cm]
z&=&s_{2j-1}z_{2j-1}&=&s_{2j}z_{2j}
\end{array}
\right.
\label{projD}
\eeq
on $\U_1,\ldots,\U_{n-3},\U_{n-1}$. The expressions
on $\U_{n-2}$ and $\U_n$ are somewhat irregular.
For later use it is enough to write the expressions of $y$ and $z$:
\beqa
&&y=\left\{
\begin{array}{ll}
z^{{n\over 2}-2}s_{n-2}t_{n-2}=s_nz^{{n\over 2}-2}&
\quad \mbox{$n$ : even}\\[0.2cm]
t_{n-2}z^{[{n\over 2}]-1}=s_nt_nz^{[{n\over 2}]-1}&
\quad\mbox{$n$ : odd}
\end{array}
\right.
\label{projDy}\\
&&z=s_{n-2}t_{n-2}z_{n-2}=s_{n}z_{n}.
\label{projDz}
\eeqa

The resolved $D_n$ surface is given by
\beqa
&&s_i+t_i^2z_i=s_i^{n-1-i}z_i^{n-i} \qquad
\mbox{in~ $\U_i$}\qquad (i\ne n-2,n)
\label{Deq1}\\
&&s_{n-2}+t_{n-2}z_{n-2}=s_{n-2}z_{n-2}^2 \qquad \mbox{in~ $\U_{n-2}$}\\
&{\rm and}&
\quad 1+s_nt_{n}^2z_{n}=z_{n}^2 \qquad\qquad \mbox{in~ $\U_{n}$}
\label{Deq3}
\eeqa
This is mapped onto the singular $D_n$ surface by (\ref{projD}), and
the map is an isomorphism except at the inverse image of the
singular point $x=y=z=0$. The inverse image consists of
$n$ rational curves $C_1,\ldots,C_n$ where $C_i$ ($i=1,\ldots,n-2$)
is the $z_i$-axis in $\U_i$ (i.e.~$s_i=t_i=0$), and also the
$t_{i+1}$-axis in $\U_{i+1}$ (i.e.~$s_{i+1}=z_{i+1}=0$).
$C_{n-1}$ and $C_n$ are the loci $t_{n-2}=z_{n-2}\mp 1=0$ parallel to
the $s_{n-2}$-axis in $\U_{n-2}$. $C_{i-1}$ and $C_i$
($i=2,\ldots,n-2$)
intersects transversely at $s_i=t_i=z_i=0$, while $C_{n-2}$
intersects also with $C_{n-1}$ and $C_n$ at
$s_{n-2}=t_{n-2}=z_{n-2}\mp 1=0$.
There is no other intersection of distinct $C_i$'s.
The self-intersection of $C_i$
in the resolved surface can be shown to be $-2$.

\begin{figure}[htb]
\begin{center}
\epsfxsize=4.5in\leavevmode\epsfbox{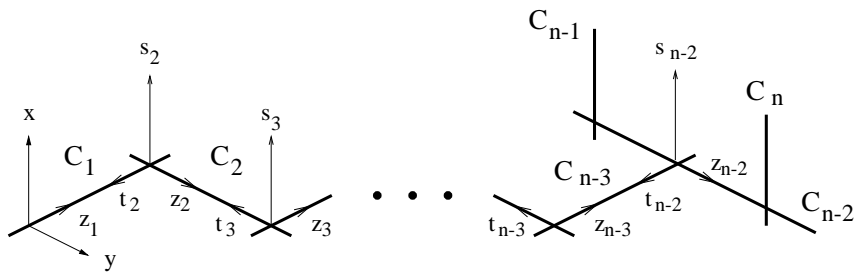}
\end{center}
\caption{resolution of $D_n$ singularity}
\end{figure}

\bigskip
\noindent
\subsection{Resolution of $E_{6,7,8}$ singularities}

\bigskip
Resolution of $E_{6,7,8}$ singularity is carried out by a
sequence of blow-ups of $\C^3$. The blow up of $x$-$y$-$z$ space
at the origin is a union of three spaces --- coordinatized by
$(x,y_1,z_1)$, $(x_2,y,z_2)$, and $(x_3,y_3,z)$ ---
which are glued together so that the
map to the $x$-$y$-$z$ space can be defined by
$(x,y,z)=(x,xy_1,xz_1)=(yx_2,y,yz_2)=(zx_3,zy_3,z)$.
In particular, there are relations $y_1x_2=1$, $z_2y_3=1$, and
$x_3z_1=1$.
The inverse image of the origin $x=y=z=0$
is a $\CP^2$. This blow-up is shown in Figure 3.

\begin{figure}[htb]
\begin{center}
\epsfxsize=4.0in\leavevmode\epsfbox{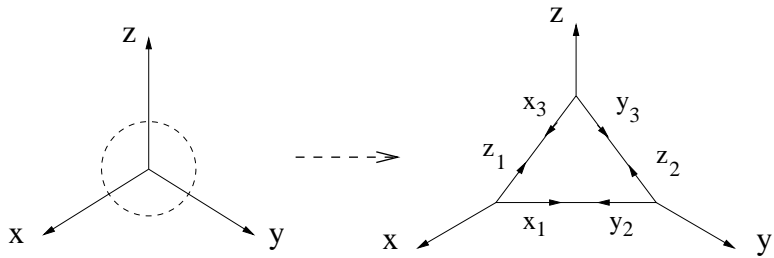}
\end{center}
\caption{blow up of the $x$-$y$-$z$ space}
\end{figure}

\bigskip
\noindent{$E_6$}

\medskip
Let us blow up the $x$-$y$-$z$ space at the origin.
The $E_6$ equation $x^2+y^3+z^4=0$ looks as
$1+xy_1^3+x^2z_1^4=0$,
$x_2^2+y+y^2z_2^4=0$, and
$x_3^2+zy_3^3+z^2=0$ in the three patches
where we ignore the $\CP^2$ described by $x=0$, $y=0$, and $z=0$
respectively.
This surface is smooth in the first two patches,
but has a singularity at the origin $x_3=y_3=z=0$ of the
third patch. In fact this is a $A_5$ type singularity
as can be seen by completing the square of $z$. The inverse image of
the singular point $x=y=z=0$ in this surface is a line $\CP^1$
defined by $x_2=y=0$ in the second patch and $x_3=z=0$
in the third patch.

Next we blow up the $x_3$-$y_3$-$z$ space at the origin.
It turns out that the surface has an $A_3$ type singularity
at one point. Continuing such process, we can finally resolve
the singularity.
The process is depicted in Figure 4. The bold lines or curves
stands for the inverse image of the singular point $x=y=z=0$.

\begin{figure}[htb]
\begin{center}
\epsfxsize=4.5in\leavevmode\epsfbox{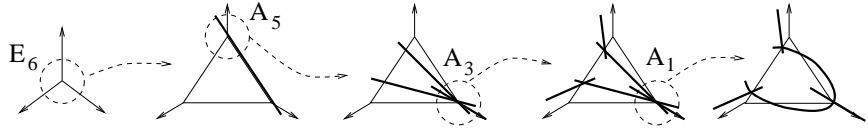}
\end{center}
\caption{the process of resolution of $E_6$ singularity}
\end{figure}

The resolved surface is defined as a hypersurface
in a 3-fold covered with five open subsets
which we denote by $\U_1,\ldots,\U_5$ and coordinatize
by $(x_1,y_1,z_1),\ldots,(x_5,y_5,z_5)$ respectively (here we have
renamed the coordinates).
These patches are glued together so that the projection to
the $x$-$y$-$z$ space is defined in the following way:
\beq
\left\{
\begin{array}{clllll}
x&=x_1y_1&=x_2y_2^6z_2&=x_3y_3^4z_3^6&=x_4y_4^2z_4^4&=x_5z_5^2\\[0.2cm]
y&=y_1&=y_2^4z_2&=y_3^3z_3^4&=y_4^2z_4^3&=y_5z_5^2\\[0.2cm]
z&=y_1z_1&=y_2^3z_2&=y_3^2z_3^3&=y_4z_4^2&=z_5
\end{array}
\right.
\label{xyz12345}
\eeq
The surface is defined by
\beqa
x_1^2+y_1+y_1^2z_1^4=0&&
\quad{\rm in}\quad \U_1,
\label{E6U1}\\
x_2^2+z_2+z_2^2=0&&
\quad{\rm in}\quad \U_2,\\
x_3^2+y_3+1=0&&
\quad{\rm in}\quad \U_3,\\
x_4^2+z_4y_4^2+1=0&&
\quad{\rm in}\quad \U_4,\\
x_5^2+z_5^2y_5^3+1=0&&
\quad{\rm in}\quad \U_5.
\label{E6U5}
\eeqa
The inverse image of the singular point
$x=y=z=0$ is a union of six rational curves
$C_1,C_2,C_{3+},C_{3-},C_{4+},C_{4-}$
which are defined in the following.
$C_1$ is the locus $x_1=y_1=0$ in $\U_1$ and $x_2=z_2=0$ in $\U_2$.
$C_2$ is the locus $y_2=x_2^2+z_2+z_2^2=0$ in $\U_2$ and
$z_3=x_3^2+y_3+1=0$ in $\U_3$.
$C_{3\pm}$ is the locus $y_3=x_3\mp i=0$ in $\U_3$ and
$z_4=x_4\mp i=0$ in $\U_4$.
$C_{4\pm}$ is the locus $y_4=x_4\mp i=0$ in $\U_4$ and
$z_5=x_5\mp i=0$ in $\U_5$.
These are depicted in the Figure 5.
One can show that any of these rational curves
has self-intersection $-2$ in the
surface. Thus, the intersection matrix is the same as the $E_6$
Cartan matrix.

\begin{figure}[htb]
\begin{center}
\epsfxsize=3.5in\leavevmode\epsfbox{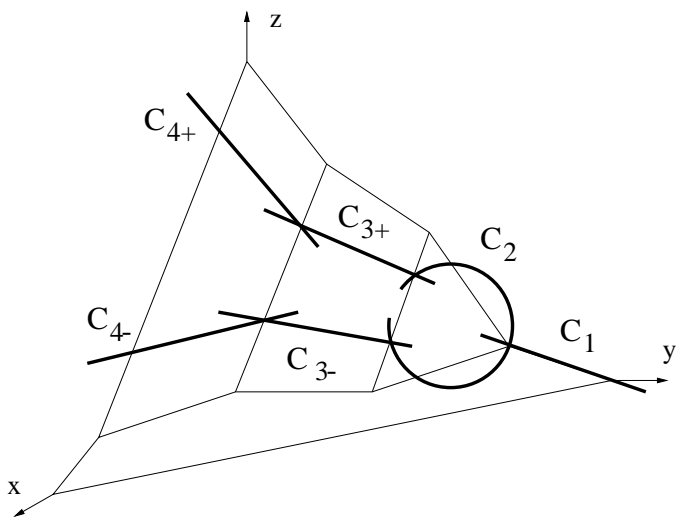}
\end{center}
\caption{resolution of $E_6$ singularity}
\end{figure}

\bigskip
\noindent{$E_7$}

\medskip
For $E_7$ we first blow up the $x$-$y$ plane at $x=y=0$.
Namely, we replace the $x$-$y$-$z$ space by
a union of $x$-$\widetilde{y}$-$z$ and $\widetilde{x}$-$y$-$z$
space that are mapped to the $x$-$y$-$z$ space by
$(x,y,z)=(x,x\widetilde{y},z)=(y\widetilde{x},y,z)$.
The equation $x^2+y^3+yz^3=0$ looks as
$x+x^2\widetilde{y}^3+\widetilde{y}z^3=0$, and
$y\widetilde{x}^2+y^2+z^3=0$ in the two patches.
The surface is smooth in the $x$-$\widetilde{y}$-$z$ space
but has a singularity at the origin
$\widetilde{x}=y=z=0$ of the second patch.
Completing the square of $y$, we have
\beq
\left(y+\frac{\widetilde{x}^2}{2}\right)^2-\frac{\widetilde{x}^4}{4}
+z^3=0.
\eeq
By putting, $x_6=y+\widetilde{x}^2/2$, $y_6=z$,
$z_6=\widetilde{x}/\sqrt{2i}$, we see that
the singularity is of the $E_6$ type
$x_{6}^2+y_{6}^3+z_{6}^4=0$.
Now, we only have to resolve this $E_6$ singularity
as done above.

The resolved $E_7$ surface is a hypersurface in a 3-fold
covered by six open subsets $\U_1,\ldots,\U_5,\U_7$
with coordinates $(x_1,y_1,z_1),\ldots,(x_5,y_5,z_5),(x_7,y_7,z_7)$.
These are glued so that the projection to
the $x$-$y$-$z$ space is defined by
\beq
\left\{
\begin{array}{cll}
x&=x_7&=\sqrt{2i}z_6(x_6-iz_6^2)\\
y&=x_7y_7&=x_6-iz_6^2\\
z&=z_7&=y_6
\end{array}
\right.
\label{xyz76}
\eeq
where $x_6,y_6,z_6$ are expressed in $\U_1,\ldots,\U_5$
as in (\ref{xyz12345}) under the replacement
$x\to x_6,y\to y_6,z\to z_6$.
We note that the coordinates of $\U_7$ and $\U_5$ are related by
$x_7=\sqrt{2i}z_5^3(x_5-i),y_7=1/(\sqrt{2i}z_5)$ and $z_7=y_5z_5^2$.

The surface is defined by
\beq
x_7+x_7^2y_7^3+y_7z_7^3=0\qquad{\rm in}\quad\U_7,
\label{E7U7}
\eeq
and by (\ref{E6U1})-(\ref{E6U5}) in $\U_1,\ldots,\U_5$.
The inverse image of the singular point $x=y=z=0$
is the union of seven rational curves $C_1,C_2,C_{3\pm},C_{4\pm}$,
$C_7$ where the first six are as given in the description of
$E_6$ surface and the last one
$C_7$ is defined by
$x_7=z_7=0$ in $\U_7$ and $x_5-i=y_5=0$ in $\U_5$.
$C_7$ intersects only with $C_{4+}$ at one point ($x_5-i=y_5=z_5=0$)
and the intersection matrix
of these curves is the same as the $E_7$ Cartan matrix.
\begin{figure}[htb]
\begin{center}
\epsfxsize=3.2in\leavevmode\epsfbox{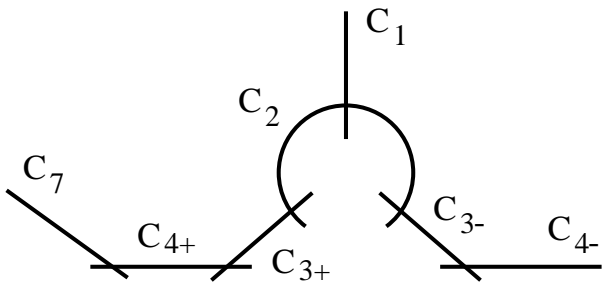}
\end{center}
\caption{resolution of $E_7$ singularity}
\end{figure}

\bigskip
\noindent{$E_8$}

\medskip
For $E_8$, we first blow up the $x$-$y$-$z$ space at the origin.
The equation $x^2+y^3+z^5=0$ looks as
$1+xy_1^3+x^3z_1^5=0, x_2^2+y+y^3z_2^5=0$ and $x_3^2+zy_3^3+z^3=0$
in the three patches. The surface is smooth in the first two patches
but has the $E_7$ singularity at the origin
$x_3=y_3=z=0$ of the third patch. Then, we just have to resolve this
$E_7$ singularity.

The resolved $E_8$ surface is a hypersurface in a 3-fold
covered by seven patches $\U_1,\ldots,\U_5,\U_7,\U_8$
with coordinates $(x_1,y_1,z_1),\ldots,(x_8,y_8,z_8)$
(here we renamed the coordinates). The 3-fold has a projection
to the $x$-$y$-$z$ space defined by
\beq
\left\{
\begin{array}{clll}
x&=x_8y_8&=x_7^2y_7&=\sqrt{2i}z_6(x_6-iz_6^2)^2\\[0.2cm]
y&=y_8&=x_7y_7z_7&=y_6(x_6-iz_6^2)\\[0.2cm]
z&=y_8z_8&=x_7y_7&=x_6-iz_6^2
\end{array}
\right.
\label{xyz876}
\eeq
where $x_6,y_6,z_6$ are expressed in $\U_1,\ldots,\U_5$
as in (\ref{xyz12345}) under the replacement
$x\to x_6,y\to y_6,z\to z_6$.
We note that the coordinates of $\U_7$ and $\U_5$ are related as
in the $E_7$ case, and the coordinates of $\U_8$ and $\U_7$
are related by $x_8=x_7/z_7,y_8=x_7y_7z_7$ and $z_8=1/z_7$.

The surface is defined by
\beq
x_8^2+y_8+y_8^3z_8^5=0\qquad{\rm in}\quad \U_8,
\label{E8U8}
\eeq
while it is defined in $\U_1,\ldots,\U_5,\U_7$
as in the $E_7$ surface. The inverse image of the singular point
$x=y=z=0$ is the union the eight rational curves
$C_1,C_2,C_{3\pm},C_{4\pm},C_7$ and $C_8$ where
$C_1$-$C_7$ are as given above in the description of
the $E_7$ surface, and $C_8$ is defined by
$x_8=y_8=0$ in $\U_8$ and $x_7=y_7=0$ in $\U_7$.
The curve $C_8$ intersects only with $C_7$ at one point
($x_7=y_7=z_7=0$). The intersection matrix of these curves is the same
as the $E_8$ Cartan matrix.
\begin{figure}[htb]
\begin{center}
\epsfxsize=3.5in\leavevmode\epsfbox{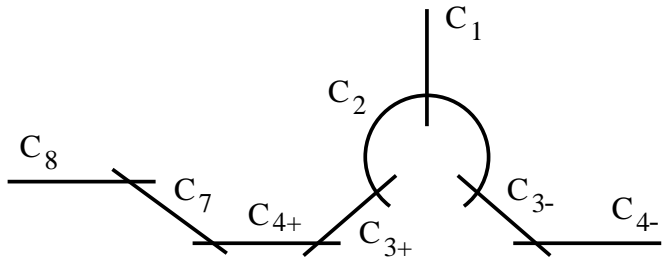}
\end{center}
\caption{resolution of $E_8$ singularity}
\end{figure}

\bigskip
\bigskip
\medskip
\noindent
\section{Geometry of Higgs Mechanism and $N=4$ Dualities in $3$
 Dimensions}

\bigskip
\medskip
In this section
we find dual pairs of three-dimensional $N=4$
supersymmetric field theories obtained by Type II string
 compactifications on Calabi-Yau 3-fold times a circle, following
the approach outlined in section 2.
We find the duals of

1'. $\prod_{i=1}^r U(k_i)$ gauge theory with $n_i$-fundamentals
for $U(k_i)$ and bi-fundamentals
$({k}_i,\overline{k}_{i+1})$

2'. $Sp(k)$ gauge theory with $n$-fundamentals.

\noindent
We also find duals of

3. Theories arising from toroidal compactification down to three dimensions
of one small $E_{6,7,8}$ instanton.  The four dimensional
versions of these theories has been considered
in \cite{Minahan, GMS, lmw}.  In particular we
prove the conjecture of \cite{intse} for the dual
of these theories.

\bigskip
The basic logic is as explained in Section 2.
We start with Type IIA string theory compactified on
$M\times S^1$ which gives the Coulomb branch of the (gauge) theory
of interest where the K\"ahler moduli of
the Calabi-Yau 3-fold $M$ corresponds to
the vector moduli of the gauge theory.
We perform the local mirror transform, obtaining a
Type IIB string theory compactified on $W\times S^1$ where
the vector moduli is now represented by the complex structure moduli
of the mirror Calabi-Yau 3-fold $W$.
Next, we consider transition to the Higgs branch corresponding
to $W^t$ through the point
where the 3-fold $W$ becomes singular.
T-dualizing on the extra circle, we can equivalently
view it as a Type IIA theory on $W^t\times S^1$ . Then, we can read
the gauge symmetry and matter content by just looking at the geometry
of the singular 3-fold, identifying the dual theory.

In this paper
we skip the first process of local mirror
transformation and refer the reader to the new paper \cite{kmv}.
Namely, we start with the Type IIB on $W\times S^1$ where
we use the result of \cite{kmv} to identify the geometry $W$
corresponding to the original gauge system (1',2',3 above).
For the cases treated in this paper, there is a simplification
 \cite{klmvw}
which is also useful.  In the cases we study here $W$ can be defined by an
equation of the form
\beq
F=uv
\eeq
where $F$ is a holomorphic function (or a section of a line bundle)
of some complex surface ${\cal S}$,
and $u$ and $v$ are complex coordinates of
another flat plane $\C^2$;
the equation defines a hypersurface (3-fold)
in the 4-fold ${\cal S}\times \C^2$.
This 3-fold can be considered as an elliptic (or $\C^*$) fibration
over ${\cal S}$
where the fibre acquires $A_{\ell-1}$ type singularity at the
zero locus of $F$. Now, we use the correspondence of Type IIB
on $A_{\ell-1}$ type singularity with Type IIA with $\ell$ NS
fivebranes \cite{OV1}. Then we can identify the Type IIB on
$W\times S^1$ as
the theory on the NS fivebrane
with worldvolume $\{F=0\}\times S^1\times\R^3$ where we note
that $\{F=0\}$ is a Riemann surface embedded in the surface ${\cal S}$.
This is the compactification on $S^1$ of the $d=4$ $N=2$ supersymmetric
gauge theory with the Seiberg-Witten curve $\{F=0\}$ \cite{Ver,klmvw}.
The results of \cite{klmvw} relating the non-compact $N=2$ curve
to the worldvolume theory of the type IIA (or equivalently M-theory)
5-brane has been recently interpreted in \cite{Witten}
as arising from the embedding of type IIA in M-theory.  This
has also been extended in \cite{Witten} to the curves
for the class of theories of the type 1' above where
the M-theory fivebrane (or equivalently type IIA fivebrane)
 is embedded in some
complex surface associated with the flavor symmetry.
In other words, we could start with the theory on such Type IIA
fivebrane\footnote{Note that the strategy we are following
in the Calabi-Yau language can be rephrased in this
case by stating that in the type IIA context compactifying
the NS 5-brane worldvolume theory on a circle and applying
T-duality on the circle we obtain NS 5-brane of type IIB,
from which we can also read off the mirror theory in 3 dimensions
by using \cite{OV1}\ or S-duality of type IIB viewing it as D5
branes.} and obtain the Type IIB geometry $W$ through the correspondence
of \cite{OV1}.
However, we stress that our main aim in this paper
is to reduce non-trivial field theory dualities to classical
symmetries of string theory.  In particular
 the Type IIB geometry $W$ can be
obtained only by knowledge of {\it classical} symmetries of
string theory \cite{kmv} (T-dualities),
without making use of non-perturbative
aspects of strings, for example how the branes of Type IIA arise
from M-theory perspective.

The starting Calabi-Yau 3-fold on which we put Type IIB string
theory to obtain the original gauge theories 1',2',3 are given by
$F=uv$ where a special case of 1' we consider separately as 1'a:

\medskip
\noindent
1'a. $U(k)$ with $n$-fundamentals\footnote{Note that if we take $k=1$ this
reduces to the abelian conifold transitions.}
\beq
F=x+z^k+y\qquad \mbox{in the $A_{n-1}$ surface
$xy=z^n$}
\label{F1}
\eeq

\medskip
\noindent
1'. $\prod_{i=1}^rU(k_i)$ with $n_i\times {k}_i$
and $({k}_i,\overline{k}_{i+1})$
\beqa
F&=&x^{r+1}+z^{k_1}x^r+\cdots
+z^{k_i+(i-1)n_1+(i-2)n_2+\cdots+n_{i-1}}x^{r-i+1}+\nonumber\\
&&\cdots
+z^{k_r+(r-1)n_1+\cdots+n_{r-1}}x+z^{rn_1+\cdots+n_r}
\label{F2}\\
&&\mbox{in the $A_{n_1+\cdots+n_r-1}$ surface
$xy=z^{n_1+\cdots+n_r}$}\nonumber
\eeqa

\medskip
\noindent
2'. $Sp(k)$ with $n$-fundamentals
\beq
F=y-z^k \qquad\mbox{in the $D_n$ surface
$x^2+y^2z=z^{n-1}$}
\label{F3}
\eeq

\medskip
\noindent
3. Critical $E_{6,7,8}$ tensionless string theories compactified
to 4 dimensions
\beqa
E_6&:& \quad F=z\qquad\mbox{in the $E_6$ surface
$x^2+y^3+z^4=0$}\\
E_7&:& \quad F=z\qquad\mbox{in the $E_7$ surface
$x^2+y^3+yz^3=0$}\\
E_8&:& \quad F=z\qquad\mbox{in the $E_8$ surface
$x^2+y^3+z^5=0$}
\eeqa

\medskip
\noindent
Some remarks are now in order.

\noindent
$\bullet$ By $ADE$ surfaces, we mean the resolved surfaces described
in the previous section.

\noindent
$\bullet$ If we consider $F=0$ as a Riemann surface factor of a IIA
theory (or equivalently
M-theory) fivebrane, the expressions (\ref{F1}) were
derived in \cite{klmvw, kkv} and the expressions (\ref{F2})
were obtained in \cite{Witten}.
This latter case
has also been recently derived using just local mirror symmetry \cite{kmv}
extending the earlier work of \cite{kkv}.
In this case, we actually
need to use $r$ different functions on $r$ different patches
of the resolved surface, each proportional to
the function $F$ in (\ref{F2}),
which also arises naturally in \cite{kmv}. (See Section 4.2 for detail)

\noindent
$\bullet$ The description (\ref{F3}) for $Sp(k)$ gauge theory
can be generalized to the case where the bare mass $m_i$ and
adjoint vev $\phi_a$ are turned on:
\beq
F=y-\prod_{a=1}^k(z-\phi_a^2)
\eeq
in the deformed $D_n$ surface (in the convention of \cite{KM})
\beq
x^2+y^2z=
{1\over z}
\left(
\prod_{i=1}^n(z+m_i^2)-\prod_{i=1}^nm_i^2\right)
-2\prod_{i=1}^nm_i\,y.
\eeq
The curve $F=0$ is exactly the same as the Seiberg-Witten
curve for $Sp(k)$ gauge theory found in \cite{AS}.
It should be
a straightforward application of the local mirror
transform to obtain the Calabi-Yau 3-fold
$F=uv$ as the Type IIB geometry for
this gauge theory.
Also, we note that this curve $F=0$ in the $D_n$ surface
can be obtained as a factor of an M-theory fivebrane
by generalizing the argument of \cite{Witten}
to the case where there is an orientifold six-plane
parallel to the D sixbranes.  Note that orientifolding
converts the $A$-singularity associated to the D6 branes
to $D$-singularity as is appearing in the above equation.

\medskip
We consider the case where there is a complete Higgs phase.
Specifically, in the class 1'a, $n\geq 2k$, in the calss 1'
$n_i+k_{i-1}+k_{i+1}\geq 2k_i$ for any $i$, in the class 2'.
$n\geq 2k+2$. This condition is equivalent to non-asymptotic free
condition of the corresponding four dimensional gauge theory.
\footnote{The $N=2$ results for Coulomb branch which we are
using also make
sense in the non-asymptotically free region.  However
there is another way to use the $N=2$ results by embedding
the non-asymptotically free theories in
asymptotically free theories in four dimensions.
For example, if we consider an
$SU(k^{\prime})$ gauge theory with flavor $n$ where
$k^{\prime}$ is chosen large enough $2k^{\prime}>n$,
there is a non-Baryonic branch of dimension $k(n-k)$ and at
a generic point of the root of that branch the theory
is identified as $U(k)$ gauge theory with $n$-flavors tensored with
free $U(1)^{k^{\prime}-k-1}$ Maxwell theory \cite{APS}.
The curve at such a point is given by
\beq
x+z^{k^{\prime}}+u_2z^{k^{\prime}-2}+
\cdots+u_{k^{\prime}-k}z^k+y=0\qquad\mbox{in the $A_{n-1}$ surface}.
\label{AF}
\eeq
Away from the $A_{n-1}$
singularity $x=y=z=0$
the curve has genus $k^{\prime}-k-1$
and this is responsible for the free Maxwell
theory part.
Thus, the behavior of the curve near $x=y=z=0$ is relevant for
the $U(k)$ gauge theory with $n$-flavors. In such a region, the higher
power $z^{k+j}$ in (\ref{AF}) is negligible compared to $z^k$.
Thus, we may well start with (\ref{F1}).
The same can be said about other cases.}

$\bullet$ The class 3 has been considered in \cite{Minahan, GMS, lmw}
and in particular local mirror symmetry applied to this problem
results in the description of the curve given above.
\bigskip
\medskip
\noindent
\subsection{$U(k)$ gauge theory with $n$-fundamentals}

\bigskip
We are interested in the locus of $F=0$ in the surface,
where the Calabi-Yau 3-fold
$W$ described by $F=uv$ acquires $A$-type singularity. Suppose
$F$ has zero of order $\ell$ along a rational curve ($\cong \CP^1$)
described by $z=0$: $F\sim z^{\ell}$. Then, $W$ has $A_{\ell-1}$-type
singularity $z^{\ell}=uv$
along the rational curve and leads to the $U(\ell)$
gauge symmetry in the Type IIA side (i.e. after T-duality on $S^1$).
When the curve $z=0$ is not a finite $\CP^1$ but has an infinitely
large volume with respect to the scale of interest, the
gauge coupling (proportional to the inverse of the volume) is
infinitesimally small compared to other couplings, and the $U(\ell)$
should be considered as a flavor symmetry.

\medskip
Let us look at $F=x+z^k+y$ in the $i$-th patch $U_i$
of the resolved $A_{n-1}$ surface which is
coordinatized by $(x_i,z_i)$. Since $(x,y,z)$ is expressed as
(\ref{projA}), $F$ is given by
\beq
F=x_i^iz_i^{i-1}+x_i^kz_i^k+x_i^{n-i}z_i^{n+1-i}.
\eeq
The locus $F=0$ looks differently depending on $i$.
We recall now that we are considering the case $n\geq 2k$.
For $1\leq i\leq k$, the first term is of lowest order both in
$x_i$ and $z_i$. For $k+1\leq i\leq n-k$ the lowest order
is the second term, and for $n-k+1\leq i\leq n$ it is the last term.
Thus, $F$ factorizes as
\beqa
&&F=x_i^iz_i^{i-1}(1+x_i^{k-i}z_i^{k+1-i}+x_i^{n-2i}z_i^{n-2i+2})
\qquad i=1,\ldots,k,\label{111}\\
&&F=x_i^kz_i^k(x_i^{i-k}z_i^{i-k-i}+1+x_i^{n-k-i}z_i^{n+1-k-i})
\qquad i=k+1,\ldots,n-k,\\
&&F=x_i^{n-i}z_i^{n+1-i}
(x_i^{2i-n}z_i^{2i-n-2}+x_i^{i-n+k}z_i^{i-n+k-1}+1)
\qquad i=n-k+1,\ldots,n.\label{333}
\eeqa
Recall that $x_i=0$ in $U_i$ and $z_{i+1}=0$ in $U_{i+1}$
defines a rational curve $C_i$. The curves
$C_{i-1}$ and $C_i$ intersect
transversely at one point $x_i=z_i=0$.
We note that the zero of the last factor in (\ref{111}) - (\ref{333})
defines a smooth curve $C$
which extends to infinity.
It intersects only with $C_k$ and $C_{n-k}$.
This can be seen by looking at the equation for $i=k,k+1$
and for $i=n-k,n-k+1$.
For example, in $U_k$, $C_k$ is given by $x_k=0$
while $C$ is given by $1+z_k+x_k^{n-2k}z_k^{n-2k+2}=0$,
and they intersect at one point
$x_k=0$, $z_k=-1$ (if $n>2k$; If $n=2k$ where $C_k=C_{n-k}$,
they intersect at two points $x_k=0$, $z_k^2+z_k+1=0$).
Likewise, it is easy to see that
$C$ and $C_{n-k}$ intersect at one point transversely.
 From the above equations we see that
$F$ has zeros at $C_i$ of order $i$ for $i=1,\ldots,k-1$,
of order $k$ for $i=k,\ldots,n-k$,
and of order $n-i$ for $i=n-k+1,\ldots,n-1$,
and also a single zero at $C$.
This is depicted in Figure 8.
\begin{figure}[htb]
\begin{center}
\epsfxsize=5.5in\leavevmode\epsfbox{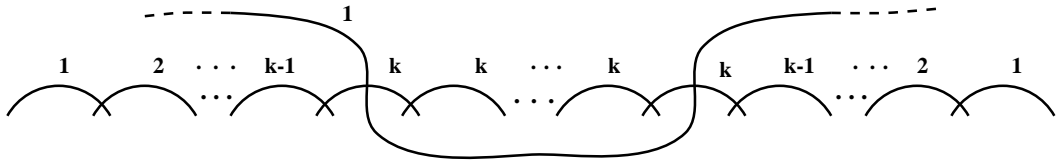}
\end{center}
\caption{the zero and the order of $F$}
\end{figure}

Now, if we look at this geometry in the Type IIA side, we see that there
is a gauge group
$U(1)\cdot U(2)\cdots U(k-1)\cdot U(k)^{n-2k+1}\cdot U(k-1)\cdots
U(2)\cdot U(1)$ coming from the $A$-type singularities
along the rational curves $C_1,\ldots,C_{n-1}$.
Note that $C$, having infinite volume
compared to others, does not lead to the gauge group.
 From the intersection of $C_i$ and $C_{i+1}$
we obtain the bi-fundamental of the neighboring gauge group, and
from the intersection of $C$ with $C_k$ and $C_{n-k}$, we obtain
fundamentals for the first and the last $U(k)$'s.
In this way, we have identified the mirror gauge theory.
\begin{figure}[htb]
\begin{center}
\epsfxsize=4in\leavevmode\epsfbox{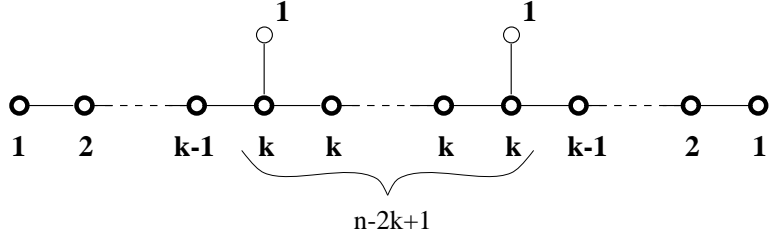}
\end{center}
\caption{the mirror of $U(k)$ gauge theory with $n$ flavors}
\end{figure}
Figure 9 depicts the quiver diagram describing the gauge and matter
content of the mirror.
To each node with index $\ell$
is associated a gauge (bold node) or flavor (normal node) group $U(\ell)$
and each edge connecting two nodes with indices $\ell_1$ and $\ell_2$
represents a hypermultiplet transforming as
$(\ell_1,\overline{\ell_2})$ under $U(\ell_1)\times U(\ell_2)$.

\bigskip
\medskip
\noindent
\subsection{Linear chain of $U(k_i)$ gauge groups}

\newcommand{\bk}{\underline{k}}
\bigskip
Next, we consider a theory with gauge group
$\prod_{i=1}^rU(k_i)$ with $n_i$-fundamentals for $U(k_i)$ and
bi-fundamentals $(k_1,\overline{k_2})$, $(k_2,\overline{k_3}),\ldots,
(k_{r-1},\overline{k_r})$.
We assume the condition
\beq
n_i+k_{i-1}+k_{i+1}\geq 2k_i
\label{Hcond}
\eeq
for the existence of a complete Higgs phase.
Before considering the mirror symmetry, we digress for a
moment to provide the precise definition of the Calabi-Yau 3-fold
$W$, or the ``function'' $F$. It turns out that we need different
functions on different patches of the resolved
$A_{n_1+\cdots+n_r-1}$ surface.
This is {\it derived} in the new paper \cite{kmv}.
Here we see how this is {\it required}
if we consider the gauge theory as coming from the worldvolume
dynamics of a fivebrane in Type IIA or M-theory.

\bigskip
\noindent
{\it Precise Definition of The Curve}

\medskip
Let us consider an $N=2$ gauge theory in four dimensions
with gauge group
$U(k_1)\times U(k_2)$ with $n_i$ massless fundamentals for $U(k_i)$
and a bi-fundamental $(k_1,\overline{k_2})$.\footnote{
We require asymptotic freedom $n_1+k_2<2k_1, n_2+k_1<2k_2$
for this part of the subsection.}
In the paper \cite{Witten}
using properties of M-theory fivebranes, it is shown that
this theory can be described by a curve\footnote{Note
that the $U(1)$'s have charged matter and thus
infrared trivial, and thus do not affect the infrared
dynamics of the non-abelian part in four dimensions.}
\beq
F:=x^3+g_1(z)x^2+g_2(z)z^{n_1}x+z^{2n_1+n_2}=0.
\label{digress}
\eeq
where $g_1(z)$ and $g_2(z)$ are polynomials in $z$ of degree $k_1$ and
$k_2$ respectively.
However, we must be careful about the precise definition of the curve
if we consider it as
embedded in the resolved $A_{n_1+n_2-1}$ surface.
At a generic point in the Coulomb branch, we expect that the theory
flows in the IR limit to a free Maxwell theory. This means
that for generic $g_1(z)$ and $g_2(z)$ the curve should
be smooth and irreducible. However, the curve (\ref{digress})
is not. $F$ is divisible by $x^2$ in the first $n_1$ patches
$U_1,\ldots,U_{n_1}$ and by $z^{n_1}x$
in the last $n_2$ patches $U_{n_1+1},\ldots,U_{n_1+n_2}$.
In order to see this, we introduce variables
$y^{\prime}=y/z^{n_2}$ and $x^{\prime}=x/z^{n_1}$.
By using the formula (\ref{projA}) for the projection,
we see that $y^{\prime}=x_i^{n_1-i}z_i^{n_1+1-i}$ and
$x^{\prime}=x_i^{i-n_1}z_i^{i-n_1-1}$ on $U_i$, and thus
that $y^{\prime}$ is well-defined on the first $n_1$ patches
while $x^{\prime}$ is defined on the last $n_2$.
By noting that $xy^{\prime}=z^{n_1}$ and $x=x^{\prime}z^{n_1}$,
$x^{\prime}y=z^{n_2}$, we see that
$F$ is divisible by $x^2$ in the first $n_1$ patches while
it is divisible by $z^{n_1}x$ in the last $n_2$ where
\beqa
&&F/x^2=x+g_1(z)+g_2(z)y^{\prime}+z^{n_2}y^{\prime 2}
\qquad\mbox{in~ $U_1,\ldots,U_{n_1}$}
\label{Fn1}
\\
&&F/(z^{n_1}x)=z^{n_1}x^{\prime 2}+g_1(z)x^{\prime}
+g_2(z)+y
\qquad\mbox{in~ $U_{n_1+1},\ldots,U_{n_1+n_2}$}
\label{Fn2}
\eeqa
Thus, the precise definition of the curve is
$F/x^2=0$ in the first $n_1$ patches and $F/(z^{n_1}x)=0$
in the last $n_2$.
This makes sense since
the two functions are related by $(1/y^{\prime})F/x^2=F/(z^{n_1}x)$
where $y^{\prime}\ne 0$ in the intersection region
because $y^{\prime}x^{\prime}=1$.

For the group $\prod_{i=1}^rU(k_i)$ with general $r$,
the function $F$ is given by
\beqa
F&=&x^{r+1}+g_1(z)x^r+\cdots
+g_i(z)z^{(i-1)n_1+(i-2)n_2+\cdots+n_{i-1}}x^{r-i+1}+\nonumber\\
&&\cdots
+g_r(z)z^{(r-1)n_1+\cdots+n_{r-1}}x+z^{rn_1+\cdots+n_r}
\label{generalr}
\eeqa
where $g_i(z)$ is a polynomial of degree $k_i$.
In the resolved $A_{n_1+\cdots+n_r-1}$ surface,
the curve is defined by
$F/x^r=0$ in the first $n_1$ patches, $F/(z^{n_1}x^{r-1})=0$
in the next $n_2$ patches, $\ldots,$
$F/(z^{(i-1)n_1+\cdots +n_{i-1}}x^{r+1-i})=0$ in the next $n_i$
patches, $\ldots$ etc.

\bigskip
\noindent
{\it Dual Gauge Theory}

\medskip
We now identify the dual gauge theory.
We present the detail for the case $r=2$.
The general case is treated in the same way.

As in the previous subsection,
it is straightforward to determine
the zero and the order of the function $F$ (\ref{F2}),
or more precisely, of $F/x^2$
(\ref{Fn1}) in the first $n_1$ patches and of $F/(z^{n_1}x)$
(\ref{Fn2}) in the remaining $n_2$ patches,
where $g_1(z)=z^{k_1}$ and $g_2(z)=z^{k_2}$.
Without loss of generality, we may assume $k_1\geq k_2$.

Using the expression (\ref{projD}),
we see that the function looks as
\beqa
&&F/x^2=x_i^iz_i^{i-1}+x_i^{k_1}z_i^{k_1}
+x_i^{n_1+k_2-i}z_i^{n_1+k_2+1-i}
+x_i^{2n_1+n_2-2i}z_i^{2n_1+n_2+2-2i}
\label{f1}
\\
&&F/(z^{n_1}x)=x_i^{2i-n_1}z_i^{2i-n_1-2}
+x_i^{i-n_1+k_1}z_i^{i-n_1+k_1-1}
+x_i^{k_2}z_i^{k_2}
+x_i^{n_1+n_2-i}z_i^{n_1+n_2+1-i}
\label{f2}
\eeqa
in the first $n_1$ and the last $n_2$ patches respectively.
As we will see below, for every $i$
there is one term $x_i^{\ell_i}z_i^{\ell_{i-1}}$ among the four
of lowest order both in $x_i$ and $z_i$. Therefore,
in each patch $U_i$, it factorizes as
\beq
x_i^{\ell_i}z_i^{\ell_{i-1}}f_i(x_i,z_i)
\eeq
where $f_i(0,0)=1$.
The curves $f_i(x_i,z_i)=0$, $i=1,\ldots, n_1+n_2$
glue up into one smooth curve $C$ that extends to infinity.\footnote{
In the special case $n_1+k_2=2k_1, n_2+k_1=2k_2$,
$f_i(x_i,z_i)$ factorizes into two, but the intersection takes place
at a point away from $x=y=z=0$ and is irrelevant for the
dynamics of interest.}
Thus, in each $U_i$ the function takes zero at $C_i$, $C_{i-1}$ and at
$C$ of order $\ell_i,\ell_{i-1}$ and $1$.

Next, we identify the lowest order term and determine $\ell_i$.
For both of the expressions (\ref{f1}) and (\ref{f2}), the following
holds:
For $i\leq k_1$ the first term is lower than the second,
for $i\leq n_1+k_2-k_1$ the second term is lower than the third,
for $i\leq n_1+n_2-k_2$ the third term is lower than the last.
The complete Higgs condition (\ref{Hcond}) yields
$k_1\leq n_1+k_2-k_1\leq n_1+n_2-k_2$.
Thus, the lowest order term is the first term
for $1\leq i\leq k_1$,
the second term
for $k_1+1\leq i\leq n_1+k_2-k_1$,
the third term for $n_1+k_2-k_1+1\leq i\leq n_1+n_2-k_2$,
and the last term
for $n_1+n_2-k_2+1\leq i\leq n_1+n_2$.
Note that $n_1+k_2-k_1\leq n_1$ by the assumption $k_1\geq k_2$.
Thus, we have
\beq
\ell_i\,=\,\left\{
\begin{array}{ll}
i&i=1,\ldots,k_1\\[0.2cm]
k_1&i=k_1+1,\ldots,n_1+k_2-k_1\\[0.2cm]
n_1+k_2-i&i=n_1+k_2-k_1+1,\ldots,{\rm min}\{n_1,n_1+n_2-k_2\}\\[0.2cm]
\left\{
\begin{array}{l}
k_2\\
2n_1+n_2-2i
\end{array}
\right.
&
\begin{array}{ll}
i=n_1+1,\ldots,n_1+n_2-k_2&{\rm if}\,\, n_1\leq n_1+n_2-k_2\\
i=n_1+n_2-k_2+1,\ldots,n_1&{\rm if}\,\,n_1>n_1+n_2-k_2
\end{array}\\[0.5cm]
n_1+n_2-i&i={\rm max}\{n_1,n_1+n_2-k_2\}+1,\ldots,n_1+n_2.
\end{array}
\right.
\eeq
The function $f_i(x_i,z_i)$ is of the following form
\beqa
f_i(x_i,z_i)=1+z_i+{\cal O}(x_iz_i),&&i=k_1,n_1+k_2-k_1,n_1+n_2-k_2
\label{three}\\
f_i(x_i,z_i)=1+x_i+{\cal O}(x_iz_i),&&i=
k_1+1,n_1+k_2-k_1+1,n_1+n_2-k_2+1\\
f_i(x_i,z_i)=1+{\cal O}(x_iz_i), &&{\rm otherwise}.
\eeqa
(Here we assume that the three values of $i$ in (\ref{three}) are
well-separated. Other case can also be treated.)
Thus, the curve $C$ intersects with $C_{k_1}$,
$C_{n_1+k_2-k_1}$ and $C_{n_1+n_2-k_2}$ transversely.

In summary, the function (\ref{f1})-(\ref{f2})
has zero at $C$ of order one and
at $C_1,C_2,\ldots,C_{n_1+n_2-1}$
of order
\beqa
&&1,2,\ldots,k_1-1,\underbrace{k_1,\ldots,k_1}_{n_1+k_2-2k_1+1},
k_1-1,\ldots,k_2+1,\underbrace{k_2,\ldots,k_2}_{n_2-k_2+1},
k_2-1,\ldots,2,1\qquad {\rm or}\nonumber\\
&&1,2,\ldots,\underbrace{k_1,\ldots,k_1}_{n_1+k_2-2k_1+1},
k_1\!-\!1,\ldots,2k_2\!-\!n_2\!+\!1,2k_2\!-\!n_2,2k_2\!-\!n_2\!-\!2,
\ldots,n_2\!+\!2,n_2,n_2\!-\!1,
\ldots,2,1\nonumber
\eeqa
if $n_2\geq k_2$, or $n_2<k_2$ respectively.
Thus, we have identified the mirror gauge theory.
The gauge group and the matter content are described by
the quiver diagram in Figure 10.
\begin{figure}[htb]
\begin{center}
\epsfxsize=4.8in\leavevmode\epsfbox{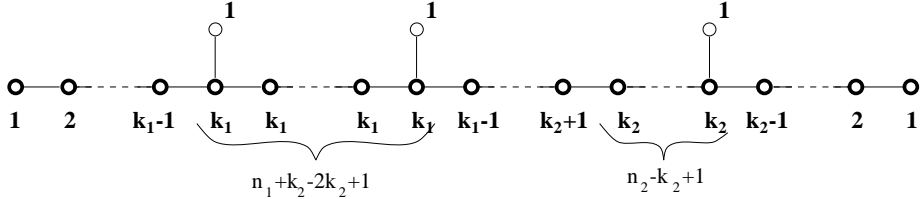}
\end{center}
\caption{the mirror in the case $n_2\geq k_2$}
\end{figure}
Here we present the mirror for the case $n_2\geq k_2$. The mirror
for the other case is obtained by
an obvious replacement of the chain of
ranks.

\bigskip
\medskip
\noindent
\subsection{$Sp(k)$ gauge theory with $n$-fundamentals}

\bigskip
In this subsection, we find the mirror of $Sp(k)$ gauge theory
with $n$-fundamental hypermultiplets. We assume $n\geq 2k+2$.

It is straightforward to determine the zero and the order of $F=y-z^k$
in the resolved $D_n$ surface.
Recall that the resolved $D_n$ surface is defined as a hypersurface
(\ref{Deq1})-(\ref{Deq3})
in
a 3-fold covered by $n$ patches ${\cal U}_1,\ldots,{\cal U}_n$.
Recall also that $y$ and $z$ are expressed
in the patch ${\cal U}_i$ by (\ref{projD}) and
(\ref{projDy})-(\ref{projDz}).
Let us look at the function $F$ in the $2j$-th patch ${\cal U}_{2j}$,
$2j\leq n-3$.
 From the expression (\ref{projD}), we see that
\beqa &&F=s_{2j}^jz_{2j}^{j-1}(1-s_{2j}^{k-j}z_{2j}^{k-j+1})\qquad
 j=1,\ldots,k-1,\\
&&F=s_{2k}^kz_{2k}^{k-1}(1-z_{2k})\qquad
 j=k,\\
&&F=s_{2j}^kz_{2j}^k(s_{2j}^{j-k}z_{2j}^{j-k-1}-1)\qquad
 j=k+1,\ldots,[(n-3)/2]
\eeqa
The last factor has a single zero at a curve $C$ which extends to
infinity. $F$ also has zeros at $s_{2j}=0$ and $z_{2j}=0$.
We now recall the defining equation of the surface
$$
s_{2j}+t_{2j}^2z_{2j}=s_{2j}^{n-1-2j}z_{2j}^{n-2j}.
$$
We see that there are two branches of zeros of $F$ for each $j$:
$s_{2j}=z_{2j}=0$ and $s_{2j}=t_{2j}=0$ which corresponds to the
rational curves $C_{2j-1}$ and $C_{2j}$ respectively.
Near the first branch $s_{2j}=z_{2j}=0$,
$(t_{2j},z_{2j})$ is a good coordinate, i.e.
$s_{2j}$ can be uniquely expressed in terms of $t_{2j}$ and
$z_{2j}$ by the defining equation.
Since $t_{2j}\ne 0$
generically, $s_{2j}\sim z_{2j}$ near $C_{2j-1}$.
Hence
$F\sim z_{2j}^j z_{2j}^{j-1}=z_{2j}^{2j-1}$ for $j\leq k$
while $F\sim z_{2j}^{2k}$ for $j>k$.
Thus, the zero of $F$ at $C_{2j-1}$ is of order $2j-1$ for
$j\leq k$ and order $2k$ for $j>k$.
Near the second branch $s_{2j}=t_{2j}=0$,
$(t_{2j},z_{2j})$ is again a good coordinate, and
$s_{2j}\sim t_{2j}^2$ for $z_{2j}\ne 0$.
Thus, $F\sim t_{2j}^{2j}$ for $j\leq k$
and $F\sim t_{2j}^{2k}$ for $j>k$ near $C_{2j}$.
Namely, $F$
has a zero at $C_{2j}$ of order $2j$ for $j\leq k$
and $2k$ for $j>k$.
By looking at the equation for $j=k$, we see that the infinite curve
$C$ and the rational curve $C_{2k}$ meet at the point
$s_{2k}=t_{2k}=0, z_{2k}=1$.
For ${\cal U}_{2j-1}$ the analysis is similar (although we need some
care for the factorization of $F$). In summary,
in the
part of the surface in the patches
${\cal U}_1,\ldots,{\cal U}_{n-3}$,
$F$ has zeros at $C_1,C_2,\ldots,C_{2k-1},C_{2k},C_{2k+1},\ldots,
C_{n-3}$ and $C$ of order $1,2,\ldots,2k-1, 2k,2k,\ldots, 2k$ and $1$
respectively.

Let us now look at the function $F$ in ${\cal U}_{n-2}$.
By looking at the expressions (\ref{projDy})-(\ref{projDz})
carefully, we see that $y$ is divisible by $z^k$,
and $y/z^k-1$ of $F=z^k(y/z^k-1)$ has a single zero at $C$.
Now, we consider the zero of
$z^k=(s_{n-2}t_{n-2}z_{n-2})^k$.
By looking at the defining equation of the surface
$$
s_{n-2}(z_{n-2}^2-1)=t_{n-2}z_{n-2}
$$
we see that there are four branches of zero:
$s_{n-2}=z_{n-2}=0$, $s_{n-2}=t_{n-2}=0$,
$t_{n-2}=z_{n-2}-1=0$ and $t_{n-2}=z_{n-2}+1=0$,
which corresponds to $C_{n-3}$, $C_{n-2}$, $C_{n-1}$ and
$C_n$ respectively.
Near $C_{n-3}$ where $s_{n-2}=z_{n-2}=0$ and $t_{n-2}\ne 0$,
the surface is coordinatized
by $(t_{n-2},z_{n-2})$
and $F\sim s_{n-2}^kz_{n-2}^k\sim z_{n-2}^{2k}$
has zero at $C_{n-3}$ of order $2k$, as we have seen.
Near $C_{n-2}$ where
$s_{n-2}=t_{n-2}=0$ and $z_{n-2}\ne 0$,
the surface is again coordinatized
by $(t_{n-2},z_{n-2})$
and $F\sim s_{n-2}^kt_{n-2}^k\sim t_{n-2}^{2k}$
has zero at $C_{n-2}$ od order $2k$.
Near $C_{n-1}$ or $C_n$ where
$t_{n-2}=z_{n-2}\mp 1=0$ and $s_{n-2}\ne 0$,
the surface is coordinatized
by $(s_{n-2},z_{n-2})$, and
$F\sim t_{n-2}^k\sim (z_{n-2}\mp 1)^k$ has zero
at $C_{n-1}$ and $C_n$ of order $k$.
The zero of $F$ in the part in ${\cal U}_{n-1}$ and ${\cal U}_n$
can be seen in the same way, but it turns out that there is no
additional zero than what we have found.

\begin{figure}[htb]
\begin{center}
\epsfxsize=5.5in\leavevmode\epsfbox{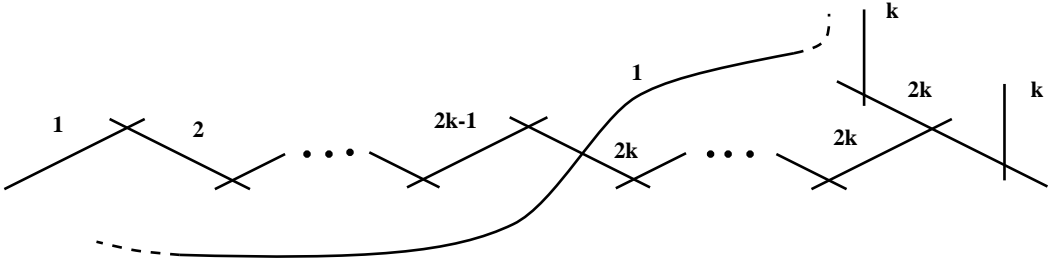}
\end{center}
\caption{the zero and the order of $F$}
\end{figure}

In summary,
$F$ has zeros at $C_1,C_2,\ldots,C_{2k-1},C_{2k},C_{2k+1},\ldots,
C_{n-3},C_{n-2},C_{n-1},C_n$ and $C$
of order $1,2,\ldots,2k-1, 2k,2k,\ldots,2k,2k,k,k$ and $1$
respectively. The curves $C_1,\ldots, C_n$ are rational curves of
finite volume whose intersection is dictated by the $D_n$ Dynkin
diagram, while $C$ extends to infinity.
The curves $C$ and $C_{2k}$ intersects transversely.
Thus, we have identified the mirror gauge theory.
It is given by the quiver diagram in Figure 12.

\begin{figure}[htb]
\begin{center}
\epsfxsize=4in\leavevmode\epsfbox{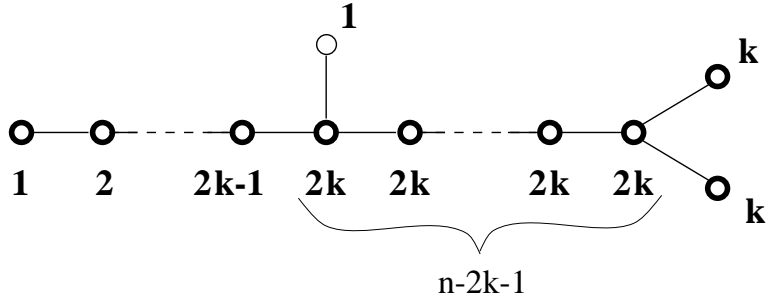}
\end{center}
\caption{the mirror of the $Sp(k)$ gauge theory with $n$ fundamentals}
\end{figure}

\bigskip
\medskip
\noindent
\subsection{Compactifications of Exceptional Tensionless String Theories}

\bigskip
In this subsection, we find mirrors of theories with
global $E_{6,7,8}$ symmetry corresponding to compactification
of theories with small $E_{6,7,8}$ instantons down to three dimensions
and show that they are ordinary gauge systems, as anticipated
in \cite{intse}.  As noted before this is a rather
interesting example in that it dualizes a gauge
system to another quantum field theory which
is expected not to have an ordinary lagrangian
description.\footnote{That the critical E-theories
do not have a lagrangian description (with finite
parameters) is strictly speaking not proven.  One
can at least rule that out as far as ordinary gauge
systems with matter are concerned.}
As in the previous cases,
we only have to determine the zero and the order of the function
$F=z$ in the resolved $E_n$ surfaces described in Section 3.3.
Thus, we follow the notation of the suitable part in that section.

\bigskip
\noindent
{\it $E_6$ Theory}

\medskip
It is a straightforward matter to see the zero and the order of $F=z$
if we look at the expression (\ref{xyz12345}) for $z$.
For example, $z=y_1z_1$ in $\U_1$, and thus it has zero at $z_1=0$
and $y_1=0$. By the defining equation of the surface (\ref{E6U1}),
the zero locus consists of the curve $z_1=x_1^2+y_1=0$
which we denote by $C$, and the locus $C_1$ of $x_1=y_1=0$.
Note that the curve $C$ extends to infinity
while $C_1$ is a rational curve with finite volume.
They intersect at one point $x_1=y_1=z_1=0$.
Near $C$ where generically $y_1\ne 0$,
the surface is coordinatized by $(x_1,z_1)$ and $z\sim z_1$ has zero at
$C$ of order 1.
Near $C_1$ where $x_1=y_1=0$ and $z_1\ne 0$,
the surface is coordinatized again by $(x_1,z_1)$
and $z\sim y_1\sim x_1^2$ has zero at $C_1$ of order 2.
The zero and the order of $z$ in
other patches $\U_2,\ldots,\U_5$
can be determined in the same way.
In summary, $F=z$ has zeros at $C_1,C_2,C_{3+},C_{3-},C_{4+},C_{4-}$
and $C$ of order $2,3,2,2,1,1$ and $1$ respectively
(see Figure 13 (a)). The curve $C$
intersects with $C_1$ at one point and extends to infinity.
The way these curves intersect is dictated by the affine $E_6$ Dynkin
diagram, where the affine node corresponds to the infinite curve $C$.
Thus, the mirror theory is a gauge theory
whose gauge and matter content is as given by
the quiver diagram in Figure 13.
\begin{figure}[htb]
\begin{center}
\epsfxsize=4in\leavevmode\epsfbox{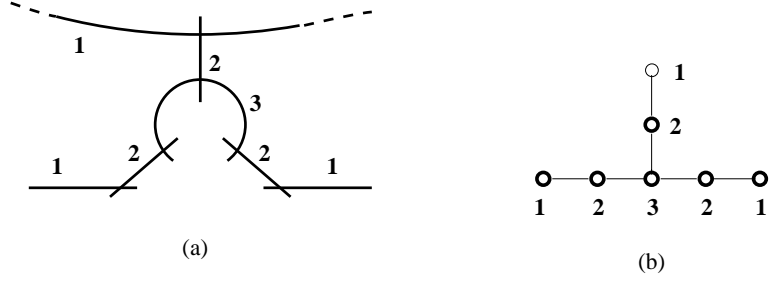}
\end{center}
\caption{(a) depicts the zero and the order of $F$ for the
$E_6$ theory.
(b) is the quiver diagram showing the mirror gauge theory.}
\end{figure}

\bigskip
\noindent
{\it $E_7$ Theory}

\medskip
The zero and the order of $F=z$ can be seen by looking at
the expression for $z$ in (\ref{xyz76}). In $\U_7$, $z=z_7$
and it has zero at $z_7=0$. By looking at the defining equation of
the surface (\ref{E7U7}), we see that there are two components:
$x_7=z_7=0$ which is the rational curve $C_7$,
and the curve $1+x_7y_7^3=z_7=0$ which we denote by $C$. The latter curve
$C$ extends to infinity and does not intersect with $C_7$.
Since $dz_1\ne 0$ in the surface near both $C$ and $C_7$,
$z=z_1$ has single zeros at $C$ and $C_7$.
In the part of the surface in the patches $\U_1,\ldots,\U_5$,
we must look at $y_6$ which is equal to ``$y$'' in the formulae
(\ref{xyz12345}) for $E_6$ case. For example in $\U_5$,
$F=z$ is given by $y_6=y_5z_5^2$. Thus, it has zero
at $z_5=0$ or $y_5=0$. By the defining equation (\ref{E6U5}),
the zero locus is $C_{4\pm}$ given by $z_5=x_5\mp i=0$ in the former
case, while it is the curves defined by $y_5=x^5\mp i=0$ in the latter
case. One of the latter curves $y_5=x^5-i=0$ is the rational curve
$C_7$. The other one $y_5=x^5+i=0$ is actually the infinite
curve $C$, as can be seen by looking at the relations
$x_7=\sqrt{2i}z_5^3(x_5-i),y_7=1/(\sqrt{2i})$.
The rational curves $C_7$ and $C_{4+}$ intersect at one point
as we have seen in Section 3.3, while
the curve $C$ intersects only with $C_{4-}$ at one point
$x_5+i=y_5=z_5=0$.
Near $C_{4\pm}$ where $z_5=x_5\mp i=0$
and $y_5\ne 0$ generically, the surface is coordinatized
by $(y_5,z_5)$ and $y_6=y_5z_5^2\sim z_5^2$ has zero at $C_{4\pm}$
of order $2$.
The zero and the order of $z=y_6$ in other patches
can be determined in the same way.
In summary, $F=z$ has zeros at $C_1,C_2,C_{3+},C_{3-},C_{4+},C_{4-},
C_7$ and $C$ of order $2,4,3,3,2,2,1$ and $1$ respectively
(see Figure 14 (a)).
The curve $C$ intersects with $C_{4-}$ at one point and extends
to infinity.
The intersection of these curves is dictated by the affine $E_7$
Dynkin diagram where the infinite curve $C$ corresponds to the
affine node.
Thus, the mirror theory is a gauge theory
whose gauge and matter content is as given by
the quiver diagram in Figure 14.
\begin{figure}[htb]
\begin{center}
\epsfxsize=4.8in\leavevmode\epsfbox{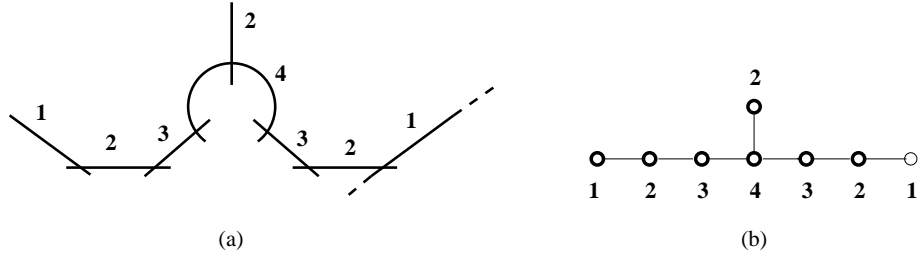}
\end{center}
\caption{(a) depicts the zero and the order of $F$ for
the $E_7$ theory.
(b) is the quiver diagram showing the mirror gauge theory.}
\end{figure}

\bigskip
\noindent
{\it $E_8$ Theory}

\medskip
For $E_8$ theory, we must look at the expression of $z$
in (\ref{xyz876}).
In the patch $\U_8$, $F=y_8z_8$ has a single zero
at the curve $C$ defined by $z_8=x_8^2+y_8=0$, and also a double
zero at the rational curve $C_8$. The curve $C$ extends to infinity
and intersects with $C_8$ at one point $x_8=y_8=z_8=0$.
The zero and the order of $F=x_7y_7=x_6-iz_6^2$
in other patches can be determined
in the same way without much effort (for the expression
of $x_6$ and $z_6$, use the formulae for $x$ and $z$
in (\ref{xyz12345})).
In summary, $F=z$ has zeros at
$C_1,C_2,C_{3+},C_{3-},C_{4+},C_{4-},C_7,C_8$ and $C$
of order $3,6,5,4,4,2,3,2$ and $1$ respectively (see Figure 15 (a)).
The intersection of these curves is given
by the affine $E_8$ Dynkin diagram where the affine node
corresponds to the infinite curve $C$.
Thus, the mirror theory is a gauge theory
whose gauge and matter content is as given by
the quiver diagram in Figure 15.
\begin{figure}[htb]
\begin{center}
\epsfxsize=5.5in\leavevmode\epsfbox{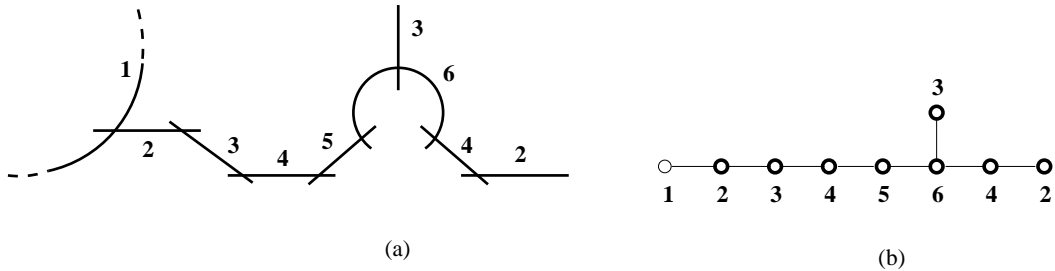}
\end{center}
\caption{(a) depicts the zero and the order of $F$ for
the $E_8$ theory.
(b) is the quiver diagram showing the mirror gauge theory.}
\end{figure}

\bigskip
\noindent
{\bf Acknowledgements}

We would like to thank S. Katz, P. Mayr, Y. Oz and A. Zaffaroni
for valuable discussions.
K.H. would like to thank
Institute for Advanced Study and
Rutgers Physics Department, H.O. would
like to thank Rutgers and Harvard Physics Departments and C.V.
would like to thank Institute for Advanced Study, for hospitality.

The researches of K.H. and H.O. are supported in part by
NSF grant PHY-95-14797 and DOE grant DE-AC03-76SF00098,
and the research of C.V. is in part supported by NSF grant PHY-92-18167.


\begin{thebibliography}{100}

\bibitem{wit} E. Witten, Nucl. Phys. B443 (1995) 85, hep-th/9503124.
\bibitem{duff} M. Duff, Nucl. Phys. B442 (1995) 47, hep-th/9501030.
\bibitem{kkv} S. Katz, A. Klemm and C. Vafa, hep-th/9609239.
\bibitem{kmv} S. Katz, P. Mayr and C. Vafa, to appear.
\bibitem{kv} S. Katz and C. Vafa, hep-th/9611090.
\bibitem{bereta} M. Bershadsky, A. Johansen, T. Pantev, V. Sadov
and C. Vafa, hep-th/9612052; C. Vafa and B. Zwiebach, hep-th/9701015.
\bibitem{ov} H. Ooguri and C. Vafa, hep-th/9702180, to appear in
Nucl. Phys. B.
\bibitem{Ah} C. Ahn and K. Oh, hep-th/9704061.
\bibitem{Ano} C. Ahn, hep-th/9705004.
\bibitem{dkv} M. Douglas, S. Katz and C. Vafa, hep-th/9609071.
\bibitem{ms} D. R. Morrison and N. Seiberg, Nucl. Phys. B483 (1997)
229, hep-th/9609070.
\bibitem{mis} K. Intriligator, D. R. Morrison and N. Seiberg, hep-th/9702198.
\bibitem{witsix} E. Witten, Nucl. Phys. B471 (1996) 121, hep-th/9603003.
\bibitem{bv} M. Bershadsky and C. Vafa, hep-th/9703167.
\bibitem{i} K. Intriligator, hep-th/9702038; J. D. Blum and
K. Intriligator, hep-th/9705030, hep-th/9705044.
\bibitem{ma} P. Aspinwall and D. R. Morrison, hep-th/9705104.
\bibitem{hw}  A. Hanany and E. Witten, hep-th/9611230.
\bibitem{dhooz} J. de Boer, K. Hori, H. Ooguri, Y. Oz and Z. Yin,
hep-th/9612131, to appear in Nucl. Phys. B.
\bibitem{kut} S. Elitzur, A. Giveon and D. Kutasov, hep-th/9702014.
\bibitem{more1} J. de Boer, K. Hori, Y. Oz and Z. Yin,
hep-th/9702154.
\bibitem{more2} J. de Boer, K. Hori and Y. Oz, hep-th/9703100.
\bibitem{more3} O. Aharony, A. Hanany, K. Intriligator, N. Seiberg,
M. J. Strassler,
hep-th/9703110.
\bibitem{more4} N. Evans, C. V. Johnson, A. D. Shapere, hep-th/9703210.
\bibitem{more5} J. H. Brodie, A. Hanany, hep-th/9704043.
\bibitem{more6} A. Brandhuber,
J. Sonnenschein, S. Theisen and S. Yankielowicz, hep-th/9704044.
\bibitem{more7} S. Elitzur, A. Giveon, E. Rabinovici,
A. Schwimmer and D. Kutasov, hep-th/9704104.
\bibitem{more8} O. Aharony and A. Hanany, hep-th/9704170.
\bibitem{OV1} H. Ooguri and C. Vafa, Nucl. Phys. B463 (1996) 55,
hep-th/9511164.
\bibitem{intse} K. Intriligator and N. Seiberg, Phys. Lett. B387
(1996) 513.
\bibitem{dhoo} J. de Boer, K. Hori, H. Ooguri and Y. Oz,
hep-th/9611063,
to appear in Nucl. Phys. B.
\bibitem{PZ} M. Porrati and A. Zaffaroni, hep-th/9611201.
\bibitem{klm} A. Klemm and P. Mayr, hep-th/9601014.
\bibitem{kamp} S. Katz, D. R. Morrison and M. R. Plesser,
Nucl. Phys. B477 (1996) 105, hep-th/9601108.
\bibitem{geosin} M. Bershadsky, K. Intriligator, S. Kachru,
D. R. Morrison,
V. Sadov, and C. Vafa, Nucl. Phys. B481 (1996) 215, hep-th/9605200.
\bibitem{mge} S. Katz and C. Vafa, hep-th/9606086.
\bibitem{morerefs} P. Berglund, S. Katz, A. Klemm, P. Mayr,
Nucl. Phys. B483 (1997) 209, hep-th/9605154.
\bibitem{bsav} M. Bershadsky, V. Sadov and C. Vafa, Nucl. Phys. B463 (1996)
398,
hep-th/9510225.
\bibitem{klmvw} A. Klemm, W. Lerche, P. Mayr, C. Vafa and N. Warner,
Nucl. Phys. B477 (1996) 746, hep-th/9604034.
\bibitem{sesh} N. Seiberg and S. Shenker, Phys. Lett. B388 (1996) 521,
hep-th/9608086.
\bibitem{str} A. Strominger, Nucl. Phys. B451 (1995) 96, hep-th/9504090.
\bibitem{gms} B. Greene, D. R. Morrison and A. Strominger, Nucl.Phys.B451
(1995) 109, hep-th/9504145.
\bibitem{yau} Articles in  ``Essays on Mirror Manifolds,''  edited
by S.-T. Yau, International
Press, 1992.
\bibitem{kont} M. Kontsevich, Proceedings of ICM, Z\"urich 1994,
alg-geom/9411018.
\bibitem{give} A. Givental, alg-geom/9603021, alg-geom/9701016.
\bibitem{syz} A. Strominger, S.-T. Yau and E. Zaslow, Nucl. Phys. B
479 (1996) 243, hep-th/9606040.
\bibitem{mor} D. R. Morrison, alg-geom/9608006.
\bibitem{gross} M. Gross and P.M.H. Wilson, alg-geom/9608004.
\bibitem{vw} C. Vafa and E. Witten, J. Geom. Phys. 15 (1995) 189,
hep-th/9409188.
\bibitem{gmv} B. Greene, D. R. Morrison and C. Vafa, Nucl. Phys. B481
(1996) 513, hep-th/9608039.
\bibitem{canHu} P. Candelas, P. S. Green  and T. H\"ubsch,
Phys. Rev. Lett. 62 (1989) 1956.
\bibitem{bj}  M. Bershadsky and A. Johansen, Nucl. Phys. B489 (1997) 122.
\bibitem{mw}  R. Friedman, J. Morgan and E. Witten, hep-th/9701162.
\bibitem{Minahan}  J.A. Minahan and D. Nemeschansky, Nucl. Phys. B489
  (1997) 24, hep-th/9608047;
Nucl. Phys. B482 (1996) 142, hep-th/9610076.
\bibitem{GMS} O.J. Ganor, D.R. Morrison and N. Seiberg,
Nucl. Phys. B487 (1997) 93,
hep-th/9610251.
\bibitem{lmw} W. Lerche, P. Mayr and N. Warner, hep-th/9612085.
\bibitem{Ver}  E. Verlinde, Nucl. Phys. B455 (1995) 211, hep-th/9506011.
\bibitem{Witten} E. Witten,  hep-th/9703166.
\bibitem{KM} S. Katz and D.R. Morrison, Journ. Alg. Geom. 1 (1992) 449.
\bibitem{AS} P.C. Argyres and A.D. Shapere,
Nucl. Phys. B461 (1996) 437, hep-th/9509175.
\bibitem{APS} P.C. Argyres, M.R. Plesser and N. Seiberg,
Nucl. Phys. B471 (1996) 159, hep-th/9603042.


\end{thebibliography}
\end{document}